\documentclass[10pt,a4paper]{article}
\usepackage[margin=0.8in]{geometry}
\usepackage{fontspec}
\usepackage{microtype}
\usepackage{graphicx}
\usepackage{booktabs}
\usepackage{array}
\usepackage{tabularx}
\usepackage{multirow}
\usepackage{caption}
\usepackage{enumitem}
\usepackage{xcolor}
\usepackage{pifont}
\usepackage{amsmath}
\usepackage{float}
\usepackage[numbers,sort&compress]{natbib}
\usepackage[hidelinks]{hyperref}

\newfontfamily\devafont[Script=Devanagari,Scale=0.92]{NotoSansDevanagari-Regular.ttf}
\newfontfamily\telufont[Script=Telugu,Scale=0.92]{NotoSansTelugu-Regular.ttf}
\newfontfamily\oriyafont[Script=Oriya,Scale=0.92]{NotoSansOriya-Regular.ttf}
\newcommand{\dev}[2]{{\devafont #1}\,(\textit{#2})}

\newcommand{\devplain}[1]{{\devafont #1}}

\setlist[itemize]{nosep,leftmargin=1.4em}
\setlist[enumerate]{nosep,leftmargin=1.6em}
\newcolumntype{L}{>{\raggedright\arraybackslash}X}
\newcolumntype{R}{>{\raggedleft\arraybackslash}p{1.6cm}}
\newcommand{\tabsize}{\footnotesize}
\newcommand{\flag}[1]{\textcolor[HTML]{A6472B}{#1}}
\newcommand{\best}[1]{\textcolor[HTML]{1B7A3D}{\textbf{#1}}}
\newcommand{\tick}{\textcolor[HTML]{1B7A3D}{\ding{51}}}
\newcommand{\cross}{\textcolor[HTML]{A6472B}{\ding{55}}}

\newcommand{\pilot}{\textsuperscript{p}}
\newcommand{\llmj}{\textsuperscript{L}}
\graphicspath{{figures/}}
\newcommand{\relcloudlo}{16\%}
\newcommand{\relcloudhi}{23\%}
\newcommand{\relondevice}{5\%}
\newcommand{\relmultilo}{16\%}
\newcommand{\relmultihi}{42\%}
\newcommand{\relmulticloudlo}{32\%}
\newcommand{\relmulticloudhi}{42\%}

\title{\textbf{Model-Agnostic and Language-Agnostic Voice Pipeline Improvement for the Agriculture Domain}}
\author{%
Aakash Singh\thanks{aakash@digitalgreen.org} \and
Lakshmi Pedapudi\thanks{Corresponding author: laxmigenius@gmail.com} \and
Chandrashekar M S\thanks{chandrashekar@digitalgreen.org} \and
Sanyam Singh\thanks{sanyam@digitalgreen.org} \and
Naga Ganesh\thanks{naga@digitalgreen.org} \and
Vineet Singh\thanks{vineet.vinsing@gmail.com}\\[4pt]
\normalsize Digital Green}
\date{}

\begin{document}
\maketitle

\begin{abstract}
FarmerChat is Digital Green's AI-powered agricultural advisory assistant for smallholder farmers,
who reach it in their own language through text, voice, or photographs, whichever is most
convenient; it has answered millions of questions for hundreds of thousands of farmers across
several countries~\citep{singh2026conversational}. Voice is a critical channel for this population,
because many users have limited literacy or limited comfort typing in their own script, so a large
share of questions arrive as field recordings made on low-end phones, often over farm machinery such
as a tractor or pump, a television or radio playing nearby, or a second person helping them use the
app. General-purpose automatic speech recognition (ASR) transcribes this audio poorly. Field noise,
extra speakers, and dense agricultural vocabulary each degrade the transcript, and the errors that
matter most fall on the crop, pest, chemical and quantity terms that carry the meaning of the
question.

This paper reports how Digital Green is addressing these failures in its own voice advisory service.
We build a modular pipeline that wraps an unmodified ASR model and targets each failure mode with a
dedicated stage: audio analysis with gated enhancement, speaker diarization with target-speaker
selection, the ASR call itself, domain-aware correction against a weighted agricultural lexicon, and
a quality gate that catches failed transcripts before they reach the downstream advisory model,
whose design is specified here and whose measurement is future work. The pipeline improves transcription without fine-tuning the ASR
model, replacing the provider, or adding a large agentic system.

We evaluate it on human-annotated FarmerChat recordings in Hindi, Telugu and Odia, scoring each
stage with word error rate and a domain-weighted error rate that penalizes agricultural-term
mistakes more heavily, rather than word error rate alone. The gain concentrates on multi-speaker
audio: once the diarizer isolates the farmer's own speech, competing voices stop entering the
transcript, and this effect holds across ASR models from different model families, which is what
makes the improvement model-agnostic. Because a general-purpose diarizer does not transfer cleanly
across these languages, the segmenter is the one component we fine-tune; every other stage uses an
off-the-shelf model behind a common interface. The design targets incremental deployment, with each
stage configurable and independently replaceable, and enhancement and correction applied
conditionally, where the evidence supports them.

On the full corpus the pipeline lowers word error rate by a relative \relcloudlo{} to \relcloudhi{}
on three cloud ASR models and \relondevice{} on an on-device model, and by \relmulticloudlo{} to
\relmulticloudhi{} on multi-speaker audio, \relmultilo{} on the on-device model. Every drop is
statistically significant on all four. Section~\ref{sec:results} reports the stage-by-stage results.

\end{abstract}

\section{Introduction}
\label{sec:intro}

\subsection{Motivation}
A farmer standing in a field can talk. Typing is much harder. So voice is the natural way in, and
the audio that arrives is nothing like clean benchmark speech:
\begin{itemize}
  \item it is recorded in the open air on a cheap phone microphone, with other people talking, farm
    machinery running, a television or radio playing, and wind;
  \item more than one person speaks, because an extension worker or a relative often helps the
    farmer use the app and talks while doing it;
  \item speakers mix languages, and they use farming words (crops, pests, chemicals, doses, units)
    that general ASR models rarely hear;
  \item nothing warns anyone when the transcript is wrong, so the advisory model answers a question
    the farmer never asked.
\end{itemize}

Not every word costs the same. Getting a function word wrong is harmless. Getting the crop, the
disease, the pest, the chemical, the quantity or the place wrong changes the question, and a changed
question can send a farmer to the wrong treatment. Table~\ref{tab:confusions} lists the confusions
we see most often in our own recordings. Plain word error rate hides them, because it charges the
same for every word.

\subsection{Problem Statement}
The production ASR model is tuned for general transcription quality. It is not tuned for getting
farm queries right. This paper asks a narrow question: can a few cheap processing stages, plus
repair of farming words, make an \emph{existing} ASR model more useful without
\begin{itemize}
  \item fine-tuning the ASR model,
  \item changing the ASR model provider,
  \item building a large agent or language-model system,
  \item or adding much compute or serving infrastructure?
\end{itemize}
The pipeline measured here meets all four conditions, and none of its stages makes a language-model
call of its own. A
small contextual pass over the cases the fixed rules cannot settle safely is named as future work
(\S\ref{sec:future}), not as part of the measured system.

\subsection{Research Questions}
\begin{itemize}
  \item \textbf{RQ1.} How much does cleaning the audio help, and for which ASR models?
    (\S\ref{sec:m0}, \S\ref{sec:res-denoise})
  \item \textbf{RQ2.} How much does splitting the speakers and keeping the farmer help when more
    than one person talks? (\S\ref{sec:fail-speakers}, \S\ref{sec:m1}, \S\ref{sec:res-ladder})
  \item \textbf{RQ3.} Can farming words be repaired without touching the ASR model?
    (\S\ref{sec:m3}, \S\ref{sec:res-correction})
  \item \textbf{RQ4.} What does all this cost in money, latency and deployment effort?
    (\S\ref{sec:alternatives}, \S\ref{sec:res-cost})
\end{itemize}
Whether the final answer to the farmer is good is outside the scope of this paper, because it depends
on the model that reads the transcript (\S\ref{sec:method-metrics}). The closest measurable thing is
how many farming terms survive, and that is reported at every stage.

\subsection{Contributions}
\begin{enumerate}
  \item \textbf{A voice pipeline that does not depend on the ASR model or the language.} Five
    stages that can each be replaced, one of them the ASR model itself, left untouched. Only the diarization
    segmenter is fine-tuned (\S\ref{sec:design}, \S\ref{sec:modules}).
  \item \textbf{A rule-based repair stage, and the word list behind it.} About 18,000 farming terms
    mined from roughly 100,000 farmer queries. Every sound-alike pair is labelled as either the same
    word twice or two different words, so the repair stage knows which pairs need context and must
    be left alone. The build runs again per language (Appendix~\ref{app:lexicon}).
  \item \textbf{A measurement of every stage} on four ASR models, cloud and on-device, in three
    languages, scored against the farmer's own words. We report word error rate, a weighted error
    rate that charges more for farming words, a farming-term rate, and paired bootstrap intervals
    (\S\ref{sec:method}, \S\ref{sec:results}). The full pipeline cuts word error rate by
    \relcloudlo{} to \relcloudhi{} on the three cloud ASR models, \relondevice{} on the on-device
    model, and \relmultilo{} to \relmultihi{} where more than one person speaks, the largest gains
    landing where there is most to fix.
  \item \textbf{Evidence for when each conditional stage pays, and the gating that follows from
    it.} Cleaning helps a generative model on its noisy tail and is not worth running on the others,
    so it is switched on per model and per noise tier. Repair is restricted to words the mining
    corpus has never seen, which is what lets it recover garbled farming terms while leaving valid
    words alone (\S\ref{sec:res-denoise}, \S\ref{sec:res-correction}).
  \item \textbf{Open-source releases.} The evaluation set, with both the farmer reference and the
    full per-speaker transcription for every
    clip,\footnote{Evaluation set (CC-BY-4.0): \url{https://huggingface.co/datasets/DigiGreen/agri-voice-eval}~\citep{digitalgreen2026evalset}}
    the word list with its labelled
    pairs,\footnote{Farming word list (CC-BY-4.0): \url{https://huggingface.co/datasets/DigiGreen/agri-lexicon-hindi}~\citep{digitalgreen2026lexicon}}
    the fine-tuned diarization
    checkpoint,\footnote{Diarization segmenter (MIT): \url{https://huggingface.co/DigiGreen/pyannote-segmentation-agri-indic}~\citep{digitalgreen2026segmenter}}
    the pipeline and evaluation
    code,\footnote{Pipeline and evaluation code: \url{https://github.com/aakashdg/agri-voice-pipeline}~\citep{digitalgreen2026code}}
    and a public
    demonstration\footnote{Demonstration: \url{https://huggingface.co/spaces/DigiGreen/farmerchat-voice-pipeline-demo}~\citep{digitalgreen2026demo}}
    (\S\ref{sec:release}).
\end{enumerate}

\section{Existing System and Baselines}
\label{sec:existing}

\subsection{Production Voice Pipeline}
\label{sec:production}
\begin{figure}[ht]
\centering
\includegraphics[width=0.94\textwidth]{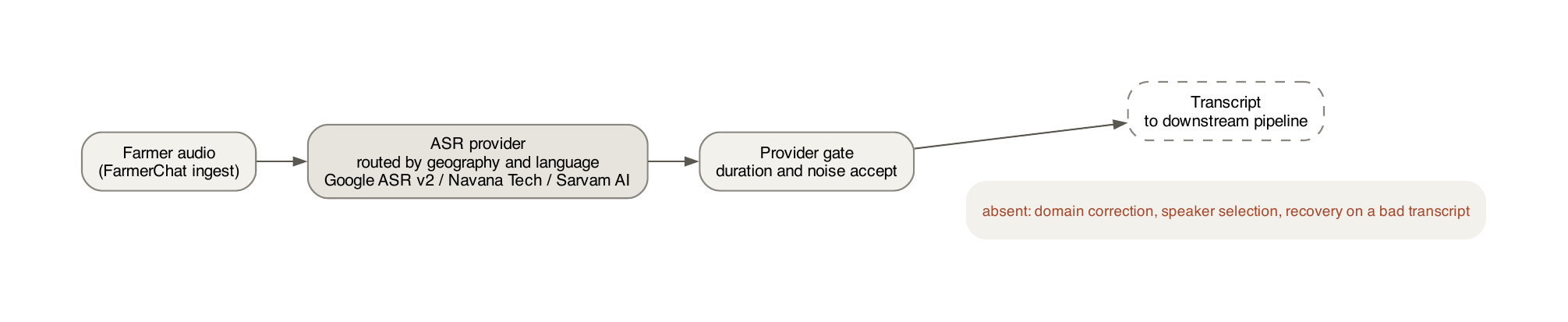}
\caption{The pipeline in production today. Red marks the three things it does not do.}
\label{fig:existing}
\end{figure}

Figure~\ref{fig:existing} shows what happens to a farmer's recording today. FarmerChat records the
question on the farmer's or the extension worker's phone. One ASR model call follows, from a
different provider per country and language: Google ASR v2, Navana Tech or Sarvam AI. The provider,
not Digital Green, then accepts or rejects the clip on length and noise. Whatever survives goes
downstream, where the rest of the advisory pipeline treats the transcript as if the farmer had typed
it. Nothing on that path looks after farming words, several speakers, or a transcript that is simply
wrong.

\subsection{Baseline Systems}
\label{sec:baselines}
Four comparison points frame this work. Every result is measured against the first; the other three
set the context, and \S\ref{sec:method-configs} says which of them was run.

\begin{itemize}
  \item \textbf{Baseline 0, vanilla ASR model.} Raw audio in, transcript out, no gate and no repair. The
    floor.
  \item \textbf{Baseline 1, production pipeline.} The current FarmerChat path of
    Figure~\ref{fig:existing}. It serves live traffic with no paired reference, so it
    sets the deployment context the pipeline has to fit.
  \item \textbf{Baseline 2, native-audio model.} One multimodal model answers straight from the audio, with
    no transcript step. It is the costlier alternative, and its answers have to be weighed against
    its cost, its latency, and the loss of a transcript anyone can score. That last point is the
    reason every configuration here produces a transcript, even when the ASR model is itself a
    multimodal model (\S\ref{sec:alternatives}).
  \item \textbf{Baseline 3, the pipeline's own ladder.} The stages of \S\ref{sec:modules} switched on
    one after another, every configuration scored against the same reference
    (\S\ref{sec:method-configs}). This is what shows what each stage adds.
\end{itemize}

\subsection{Related Work}
\label{sec:related}
Our earlier benchmark~\citep{chandrashekar2026asr} tested ten ASR models on a multi-language corpus
of farmer recordings (Table~\ref{tab:datasets}). It introduced the Agriculture Weighted Word Error
Rate (AWWER), and it showed that picking the main speaker after diarization is the single most
useful thing you can do to multi-speaker audio. This paper builds the pipeline that benchmark asked
for and measures it as a whole. The advisory model that reads these transcripts, and the way its
answers are checked fact by fact, are described in \citet{singh2026conversational}.

General speech recognition has improved quickly, both in large multilingual models such as
Whisper~\citep{radford2023whisper} and Meta's Massively Multilingual Speech
project~\citep{pratap2023mms} and in commercial cloud services. Indian-language coverage has grown
through national and academic work: Bhashini~\citep{bhashini2022}, Vaani~\citep{vaani2023}, work
toward recognition for the next billion users~\citep{javed2022nextbillion},
Vakyansh~\citep{chadha2022vakyansh} and SPRING-INX~\citep{nithya2023springinx}. Most of that
training audio is read or broadcast speech in fairly clean conditions, which is not how farmers
speak, and coverage across our three languages is still uneven~\citep[Table~II]{chandrashekar2026asr}.
That is why we treat the ASR model as a part to be swapped rather than a thing to be fixed.

The processing stages use established tools: DeepFilterNet for
cleaning~\citep{schroter2023deepfilternet}; pyannote~\citep{bredin2023pyannote,plaquet2023powerset},
ECAPA-TDNN embeddings~\citep{desplanques2020ecapa}, NeMo MSDD and
Sortformer~\citep{park2022msdd,park2024sortformer} and DiariZen~\citep{han2025diarizen} for
diarization; Silero for voice activity detection~\citep{silero2024vad}. Diarization is scored with
DER and with concatenated minimum-permutation WER (cpWER) as defined for
CHiME-6~\citep{watanabe2020chime6}. For repairing domain words after recognition,
\citet{garg2020entitycorrector} combine spelling distance, sound similarity and context to keep only
plausible candidates, and \citet{ma2024asrcorrection} restrict a corrector to a fixed candidate list
instead of letting it rewrite freely. Our repair stage keeps the first two signals and the fixed
list, and leaves anything that needs context alone (\S\ref{sec:m3-algo}).

\section{Failure Analysis of the Existing Pipeline}
\label{sec:failure}
This section is about where Baseline 0 and Baseline 1 break. The evidence comes from the published benchmark corpus
(Table~\ref{tab:datasets}; recorded June 2024 to February 2025)~\citep{chandrashekar2026asr}, the
human quality-checked corpus, and the reference annotated corpus. Pilot values carry the marker
\pilot{}. Values a language-model classifier produced, rather than code, carry \llmj{}. Each
subsection below takes one failure and the evidence for it.

\subsection{Domain-Specific Recognition Errors}
\label{sec:fail-domain}
\begin{itemize}
  \item Farming words get swapped for something that sounds like them, or dropped outright. Crop
    names and fertilizer or chemical names are the biggest group in all three languages. The most
    common pairs in our own recordings are in Table~\ref{tab:confusions}.
  \item Under AWWER, which charges more for farming words, the errors sit on exactly those
    words~\citep[\S4.4.4]{chandrashekar2026asr}.
  \item Sound distance does not tell you which pairs are dangerous. A word swapped for a sound-alike
    with a different meaning nearly always changes the question. The same word spelled two ways
    nearly never does. Which of the two a pair is has to be decided pair by pair, and that decision
    is what the repair stage is built on (Table~\ref{tab:pairlabels}).
\end{itemize}

\begin{table}[ht]
\centering\tabsize
\caption{The farming confusions we see most often, and what each one does to the question.}
\label{tab:confusions}
\begin{tabularx}{\textwidth}{@{}l l L r@{}}
\toprule
Reference term & ASR model output & What changes & Count \\
\midrule
\dev{दवा}{davā}, medicine or pesticide & \dev{दावा}{dāvā} & A treatment becomes a claim & 335 \\
\dev{खाद}{khād}, fertilizer & \dev{खाद्य}{khādya} & Fertilizer becomes food & 194 \\
\dev{बीज}{bīj}, seed & \dev{बीच}{bīch} & Seed becomes in-between & 153 \\
\dev{पत्ता}{pattā}, leaf & \dev{पता}{patā} & A leaf becomes an address & 144 \\
\dev{रोपने}{ropne}, to transplant & \dev{रोकने}{rokne} & Plant it becomes stop it & 88 \\
\dev{गेहूं}{gehūn}, wheat & \dev{गेम}{gem} & A crop becomes a non-word & 81 \\
\dev{मूंग}{mūng}, green gram & \dev{मुंह}{munh} & A crop becomes a body part & 74 \\
\bottomrule
\end{tabularx}
\end{table}

\subsection{Field Recording Conditions}
\label{sec:fail-noise}
\begin{figure}[ht]
\centering
\includegraphics[width=0.76\textwidth]{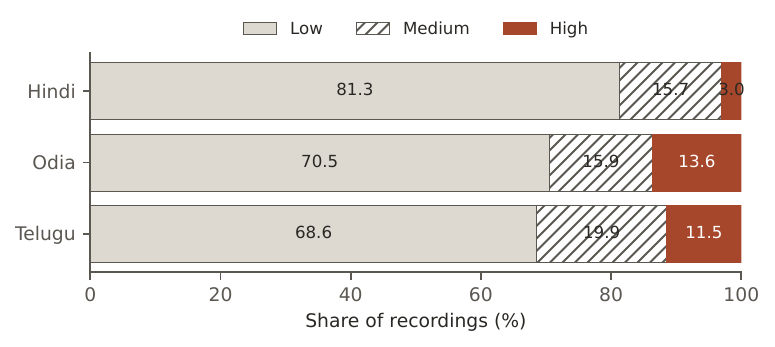}
\caption{Noise level by language, as a share of recordings~\citep[Table~I]{chandrashekar2026asr}.}
\label{fig:noise}
\end{figure}
\begin{itemize}
  \item Recordings are made outdoors on cheap phone microphones. Tractors and irrigation pumps run
    nearby. A television or radio plays. Wind hits the microphone, and the open field adds echo. Other
    people talking is the most common problem of all, in every one of the three
    languages~\citep{chandrashekar2026asr}.
  \item This is a well known weak spot. ASR models trained on clean read speech have always
    struggled with noisy, distant, echoing audio, which is the whole point of the CHiME challenge
    series~\citep{watanabe2020chime6}.
  \item The three languages are not affected equally. High-noise audio is about four times as common
    in Odia as in Hindi, with Telugu in between (Figure~\ref{fig:noise}). So a cleaning stage has to
    react to the clip in front of it instead of treating every clip the same (\S\ref{sec:m0}).
  \item Noise costs meaning, not just words. In the reference annotated corpus, the noisier the
    clip, the more often the farmer's point is lost altogether (Table~\ref{tab:intentnoise})\llmj.
\end{itemize}

\begin{table}[ht]
\centering\tabsize
\caption{How often the farmer's point survives by noise tier, one cloud ASR model against the
human reference. Language-model judged\llmj.}
\label{tab:intentnoise}
\begin{tabular}{@{}lrrr@{}}
\toprule
Noise tier & Intent preserved & Partly preserved & Intent lost \\
\midrule
Low & 83\% & 11\% & 6\% \\
Medium & 82\% & 11\% & 7\% \\
High & 78\% & 12\% & \flag{10\%} \\
\bottomrule
\end{tabular}
\end{table}

\subsection{Multiple Speakers}
\label{sec:fail-speakers}
A field visit is rarely one person talking. An extension worker or a family member often works the
app for the farmer and talks while doing it, and other people are usually around.

\begin{itemize}
  \item About one clip in five in the human quality-checked corpus has more than one speaker, and about 4\% have three or more. 
  \item Two people talking at once is one of the hardest things in recognition. When the second
    voice is nearly as loud as the farmer's, an ASR model that returns one transcript writes down
    both and cannot say which words were the question. This is why the CHiME series added
    multi-speaker tracks~\citep{watanabe2020chime6}.
  \item Multi-speaker clips are harder for every diarizer we tested, open or cloud, which is why one
    of them was fine-tuned (\S\ref{sec:m1}).
  \item Picking out the main speaker before transcription is the largest single gain available.
    Our earlier benchmark showed it across ten ASR models and three languages: split the recording
    by speaker, score only the main speaker's turns, and word error rate on multi-speaker audio
    drops a long way. The size of the drop follows how often that ASR model meets a second
    speaker~\citep[\S4.6, Table~VII]{chandrashekar2026asr}. This pipeline turns that result into a
    stage, and \S\ref{sec:res-ladder} measures what it is worth on our corpus with four ASR models
    in the identical pipeline.
  \item Without selection, the farmer's question sits buried inside the extension worker's speech.
\end{itemize}

\subsection{Lack of Error Recovery}
\label{sec:fail-recovery}
The production path has no idea whether a transcript is any good. A garbled transcript goes to the
advisory model exactly as a clean one does. There is no accept step, no reject step and no reroute
(Figure~\ref{fig:existing}). A language model answers fluently from whatever text it gets, so a
wrong transcript does not look wrong: the farmer gets a confident answer to a question nobody asked,
with nothing to warn that the input was broken. That silence makes every other failure worse, because
a noise or speaker error that gets this far is acted on instead of caught. How often a question
reaches the advisory model with its point already lost is in Table~\ref{tab:intentnoise}\llmj.

\subsection{Evaluation Blind Spot}
\label{sec:fail-eval}
The last failure is in how we measure, not in the audio.
\begin{itemize}
  \item Word-level metrics treat every substitution alike, so a ruined crop name looks the same as a
    ruined function word. This weakness of string-edit metrics is
    old news~\citep{morris2004wer}.
  \item The ASR model with the best WER is not the one that gets farming words right. On the
    published benchmark in Hindi, one system is first on WER and fifth on AWWER, while another with
    main-speaker selection is third on WER and first on AWWER~\citep[\S4.5]{chandrashekar2026asr}.
    A metric that cannot tell a usable transcript from a dangerous one cannot steer the system.
  \item Not every mangled word matters equally. Some change the question and some are still clear
    from context, and you cannot tell which from sound alone (Table~\ref{tab:pairlabels}).
  \item That is why the evaluation reports a weighted rate and a farming-term rate next to WER
    (\S\ref{sec:method-metrics}) instead of WER by itself.
\end{itemize}

\subsection{Reference Annotation Quality}
\label{sec:fail-annotation}
The last failure is in the reference itself. A person transcribing field audio hears through the
mess without meaning to, and a machine does not, so a single-line reference collected without an
explicit protocol records something slightly different from what was said. Two effects matter for
scoring, and the protocol below answers both.

\begin{itemize}
  \item \textbf{People write the word they know was meant.} An annotator who knows the crop hears
    the intended word and writes that, not the sound the farmer actually produced. Our own sheet
    shows the reflex directly: on many clips the annotator wrote the standard spelling in braces
    next to what was said (Table~\ref{tab:annotaudit}). That is useful only because it was written
    down. Without that convention the reference records the intended word rather than the spoken
    one.
  \item \textbf{One line of text hides several speakers.} A reference kept as a single sentence per
    clip merges whoever was talking. On a sizeable share of the multi-speaker clips, the single-line
    reference carries a second speaker's words as well as the farmer's. Score against that and a
    ASR model earns credit for transcribing a bystander, which is the opposite of what the service
    needs.
  \item \textbf{Fields that contradict each other.} A handful of clips carry a segment that ends
    before it starts, both speaker-count ticks at once, or an empty farming-word field that could
    mean either no farming word or no entry. Each is small, and together they are why the protocol
    fixes the schema as well as the wording.
\end{itemize}

\begin{table}[ht]
\centering\tabsize
\caption{An audit of the annotation sheet, 825 clips, 275 per language. Each row counts clips.}
\label{tab:annotaudit}
\begin{tabularx}{\textwidth}{@{}L r L@{}}
\toprule
What the audit looked for & Clips & Why it matters \\
\midrule
Standard spelling supplied in braces beside what was said & 183 & The annotator heard the intended word, not the spoken one \\
Single-line reference carries another speaker's words, of 157 multi-speaker clips & 47 & Scoring against it rewards transcribing a bystander \\
No farming word listed & 78 & Really absent, or an unfilled field; the sheet cannot say \\
Segment ends before it starts & 6 & Breaks any script that trusts the timestamps \\
Ticked single-speaker and multiple-speaker at once & 3 & The audio-issue list is not mutually exclusive \\
Speaker count disagrees with the turn labels used & 1 & Two fields that should agree, do not \\
\bottomrule
\end{tabularx}
\end{table}

A reference of this kind needs an explicit protocol. We wrote one that asks for each speaker's
turns with their own timestamps, a label naming the speaker who asked the farming question, audio
issues from a fixed list, and the farming words with their categories. What was said gets written,
and the standard form goes in braces beside it. We ran that protocol as a trial with the Indian
Institute of Science, Bangalore, and \S\ref{sec:method-annotation} sets it out with a worked clip.

\section{Design Principles and Architecture}
\label{sec:design}

\subsection{Design Principles}
The failures in \S\ref{sec:failure} lead to five rules that shaped every choice here. Put together
models that are each already good at one job. Keep every stage switchable. Judge the system by
whether the farmer's question survives, not by word-for-word match alone.

\begin{itemize}
  \item \textbf{P1, use specialists.} The ASR model, the voice activity detector, the diarizer and
    the cleaner each do the one thing they are good at (\S\ref{sec:alternatives}).
  \item \textbf{P2, retrain as little as possible.} Improve the system around the ASR model.
    Fine-tune only where nothing off the shelf works, which here is one model, the diarization
    segmenter (\S\ref{sec:m1}).
  \item \textbf{P3, make every stage configurable.} Model choice, thresholds, confidence levels and
    accept or reject rules are all settings, not code. This is what supplies the missing recovery
    step of \S\ref{sec:fail-recovery}.
  \item \textbf{P4, score the question, not the string.} Check whether the farmer's question came
    through, not only whether the words match (\S\ref{sec:fail-eval}).
  \item \textbf{P5, count cost as well as accuracy.} Accuracy, money, latency and deployment effort
    together.
\end{itemize}

Replacing or fine-tuning the ASR model instead (against P2) buys accuracy but costs provider
freedom, training and serving machinery, and a great deal of hand annotation that is itself
unreliable on noisy multi-speaker audio (\S\ref{sec:alternatives}).

\subsection{High-Level Architecture}
\label{sec:arch}
\begin{figure}[ht]
\centering
\includegraphics[width=0.76\textwidth]{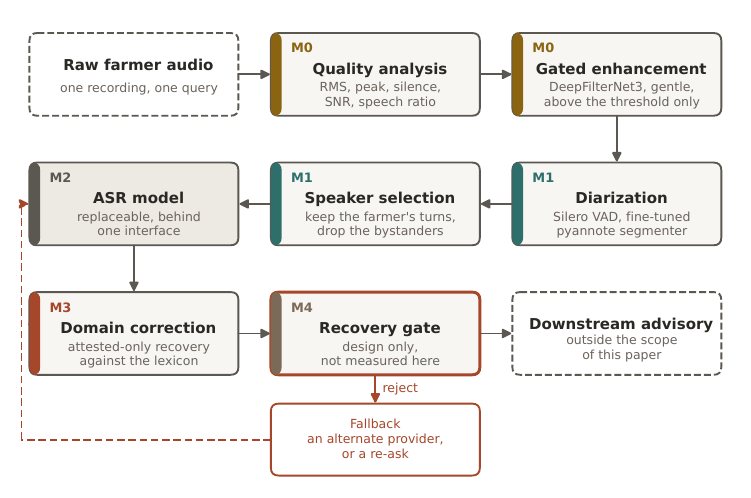}
\caption{The proposed pipeline. Coloured edges mark the modules; M4 is design only.}
\label{fig:pipeline}
\end{figure}

Five modules make up the pipeline, and one of them is the ASR model call, which can be swapped
(Figure~\ref{fig:pipeline}). Each
module can be replaced or switched off by itself, and the ladder of \S\ref{sec:method-configs}
switches them on one at a time. Table~\ref{tab:modules} says what each one does and which of
them carries a trained weight. Only the recovery gate (M4) branches. Everything else is a straight
line.

\begin{table}[ht]
\centering\tabsize
\caption{What each module does. ``Trained'' means a model weight changed for this system.}
\label{tab:modules}
\begin{tabularx}{\textwidth}{@{}l L L l l@{}}
\toprule
Module & Function & Models & Trained & Section \\
\midrule
M0 & Measures audio quality, then cleans it only if needed & Signal statistics; DeepFilterNet3 & No & \S\ref{sec:m0} \\
M1 & Finds speech, splits it by speaker, keeps the farmer & Silero VAD; pyannote segmenter and clusterer & Segmenter only & \S\ref{sec:m1} \\
M2 & Turns audio into text & Any provider behind one interface & No & \S\ref{sec:m2} \\
M3 & Repairs garbled farming words & Labelled word list; sound matcher with safety gates & No & \S\ref{sec:m3} \\
M4 & Judges the transcript and reroutes bad ones, design only & Confidence and consistency checks; routing rules & No & \S\ref{sec:m4} \\
\bottomrule
\end{tabularx}
\end{table}

\subsection{Alternative Design Strategies}
\label{sec:alternatives}
Two other routes were considered and dropped as the main plan.

\begin{itemize}
  \item \textbf{Fine-tune the ASR model.} It would learn farm speech directly, so domain words may
    come out better. But ASR models are trained on clean audio, field audio has to be transcribed by
    hand first, and on noisy clips those hand transcripts are themselves unreliable, so the model
    learns from bad references. It also needs noisy training data, training and serving machinery, and
    it locks the system to one provider.
  \item \textbf{Hand the whole job to a language model or agent.} Flexible reasoning, repair from
    context, simple to orchestrate. But the speech problems underneath, noise and several speakers,
    are still there. There is less control over the specialist models, repair quality rides on the
    prompt, and cost and latency per query go up and get harder to predict. The audio-in, answer-out
    variant leaves no transcript to score (Baseline 2, \S\ref{sec:baselines}).
  \item \textbf{The modular pipeline, which is what we built.} One specialist per job, explicit
    control at every stage, parts that swap out, measurement per part, almost no new infrastructure.
    The price is more moving parts than a single model call, and correctness that depends on tuning
    each gate.
\end{itemize}

No stage of the pipeline reported here adds a language-model call of its own. Repair follows fixed
rules, and the
contextual pass that would settle the sound-alike cases is named as future work rather than built in
(\S\ref{sec:m3-algo}, \S\ref{sec:future}). The native-audio point matters for measurement as much as
for cost. A model that goes straight from audio to an answer can only be judged on the answer: WER,
AWWER and cpWER stop being defined. Keeping a transcript in the middle is what lets every stage in
this paper be measured on its own.

\section{Pipeline Modules}
\label{sec:modules}

\subsection{M0: Audio Analysis and Enhancement}
\label{sec:m0}
The stage looks at the audio first and cleans it only where cleaning is likely to help.

\begin{itemize}
  \item \textbf{Quality detection.} RMS level, peak level, how much of the clip is silence, an
    estimate of signal against noise, and how much of the clip is speech. All of it is arithmetic on
    the waveform, computed before any model runs, so deciding whether to clean costs almost nothing.
    A single loudness threshold recovered almost none of the gain, so the gate uses all the features
    together. That gate is what a deployment runs. In the runs reported here the tier comes
    from the corpus's own noise label\llmj{}, which keeps the measured effect of cleaning separate
    from the accuracy of the gate that selects it.
  \item \textbf{Enhancement.} DeepFilterNet3~\citep{schroter2023deepfilternet}, a small CPU-only
    cleaning network, run gently instead of at its default maximum. At maximum it over-processes the
    signal and strips cues the ASR model needs, so the attenuation is capped at 6~dB. 
    On noisy clips a generative ASR model invents words that were never spoken, and gentle cleaning
    cuts that down. What the stage costs to run is in \S\ref{sec:res-cost}.
  \item \textbf{Adaptive gating.} Two ways to decide, both deployment settings. The safeguard always runs the cleaner and
    keeps its output only where it removed enough energy to matter. The compute saver lets the
    signal features decide whether to run the cleaner at all, which skips most traffic because most
    traffic is already clean. The operating point and the per-query cost are in
    \S\ref{sec:res-cost}.
  \item \textbf{Enhancement is model-dependent, so knowing when to apply it is the finding.}
    Suppression tuned to sound better
    to a human can remove information an ASR model relies on, so less measured noise does not mean
    fewer errors. The stage therefore fires only above the noise threshold and only for a model
    whose own evidence supports it, and that threshold is checked
    again whenever the ASR model changes. Cleaning everything would not have that property.
    \S\ref{sec:res-denoise} reports which of the four ASR models it pays on.
  \item \textbf{Why M0 stays in the pipeline.} It pays on the noisy tail for the ASR model it
    suits. And it also cleans the audio the diarizer then has to split, so part of its value shows up inside M1's
    result, which is why the stage is judged by noise tier rather than by a flat corpus number.
\end{itemize}

\subsection{M1: Speaker Detection and Selection}
\label{sec:m1}
Getting to the farmer takes three steps: find the speech, work out who spoke when, keep the farmer's
turns.

\begin{itemize}
  \item \textbf{Voice activity detection.} Silero VAD~\citep{silero2024vad}, cross-checked against
    the diarizer's own segmentation. Gating pyannote with Silero lowers DER at the strictest setting
    but leaves the ASR model with too few words, so the DER-minimizing setting is not the one we
    use (\S\ref{sec:res-diar}).
  \item \textbf{Diarizer comparison.} Nine systems, seven open and two cloud, scored with one
    harness, and the fine-tune of the next bullet makes ten. No stock system split Hindi, Telugu and
    Odia field audio cleanly, and most of the error is false alarm: speech heard in background
    noise. That is the error the fine-tune goes after, and \S\ref{sec:res-diar} has the comparison.
  \item \textbf{The one fine-tune in this system.} pyannote's
    segmenter~\citep{bredin2023pyannote,plaquet2023powerset}, trained on the project's own field
    audio, because no off-the-shelf model gave the false-alarm profile selection needs. It is
    scored on a frozen held-out split it never trained on, against the same reference at the same
    collar (\S\ref{sec:res-diar}). Swapping the clusterer underneath it changes almost nothing, so
    the segmenter is the lever, not the clusterer. The checkpoint and its recipe are
    released~\citep{digitalgreen2026segmenter}.
  \item \textbf{Best-speaker selection.} The rule that ships uses loudness with a duration floor.
    Among speakers who hold the floor for at least $\max(1.0~\mathrm{s},\ \tfrac{1}{4}$ of the
    longest speaker's total time$)$, the loudest on average is taken to be the farmer, because the
    farmer holds the phone. Their turns are padded a little at each end, so the first and last words
    are not clipped, and joined for the ASR model. Selection reads whatever the cleaning gate passed
    on, so the ladder measures the two stages in the order a deployment runs them. There is no voice sample of the farmer to enrol,
    so target-speaker selection from an enrolment is out of scope. 
  \item \textbf{Why loudness, and not duration alone.} A rule that picks whoever talks longest
    fails on the case that matters most: a bystander who simply holds the floor longer than the
    farmer. Loudness separates that case, because a bystander is usually further from the
    microphone. Four text-only rules were scored against the annotated main-speaker label, and the
    shipped rule end to end through the transcript it produces, both in \S\ref{sec:res-diar}.
\end{itemize}

\subsection{M2: ASR model, Unchanged and Replaceable}
\label{sec:m2}
\begin{itemize}
  \item The ASR model is the one part the pipeline does not touch. It sits behind a shared
    interface as a block you can swap: audio in, transcript out. That is what makes the system
    model-agnostic.
  \item \textbf{No single ASR model is good at all three languages.} Ten providers have been
    benchmarked on this audio and their coverage is uneven, because a system that leads one of the
    three can fall apart in another~\citep[Table~II]{chandrashekar2026asr}. So routing is per
    language rather than one model for everything, and three providers run in production today:
    Google ASR v2, Navana Tech and Sarvam AI.
  \item \textbf{ASR models in the staged evaluation.} Four systems from different families and
    deployment styles, every one run through every stage (\S\ref{sec:method-runplan}). That is what
    the model-agnostic claim is measured on.
\end{itemize}

\subsection{M3: Domain-Aware Correction}
\label{sec:m3}
The repair stage fixes farming words the ASR model got wrong. It works against a fixed, checked
word list instead of rewriting freely, and the stage makes no language-model call of its own.

The open question was whether to use a language model or fixed rules, and measurement settled it
(\S\ref{sec:res-correction}). What shipped is neither of the two original options: one fixed-rule stage
that aims at precision rather than coverage.

\subsubsection{Lexicon Construction}
\label{sec:m3-lexicon}
\begin{figure}[ht]
\centering
\includegraphics[width=0.86\textwidth]{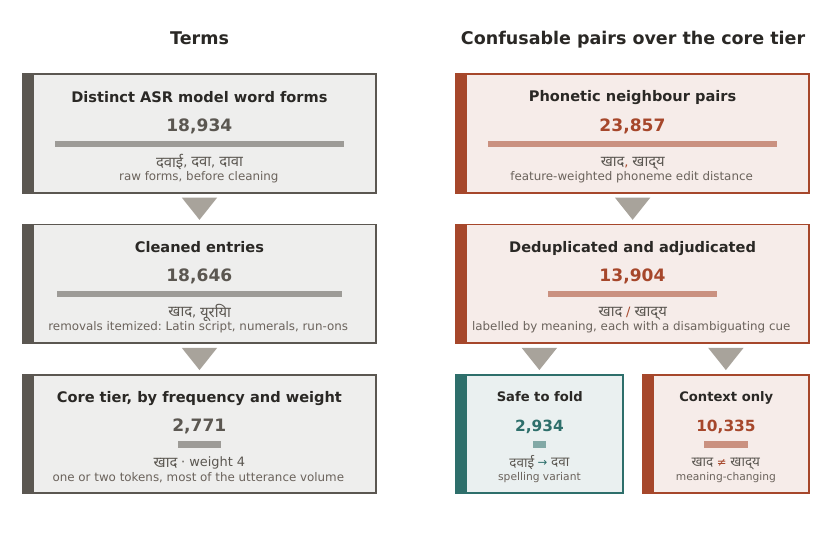}
\caption{The word list build, with one worked example at each stage. Step detail in
Appendix~\ref{app:lexicon}.}
\label{fig:lexfunnel}
\end{figure}

The word list is a contribution in itself. Farming words are mined from the reference annotated
corpus, cleaned up, grouped by how they sound, and then every sound-alike pair is labelled as either
the same word twice or two different words (Figure~\ref{fig:lexfunnel}). The full seven-step build
is Appendix~\ref{app:lexicon}, and Table~\ref{tab:pairlabels} is the labelling the repair rule
depends on. The list is released with its labelled pairs~\citep{digitalgreen2026lexicon}.

\begin{figure}[ht]
\centering
\includegraphics[width=0.85\textwidth]{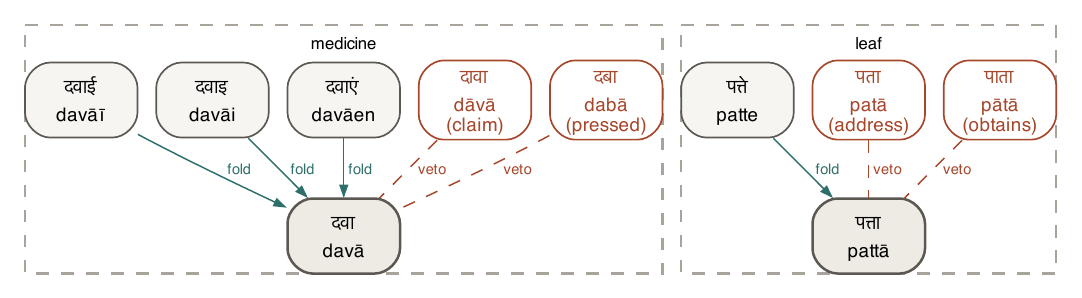}
\caption{Two variant families. Folded variants collapse into the main word; refused ones stay
separate.}
\label{fig:families}
\end{figure}

That labelling is what lets the stage tell a harmless spelling variant from a dangerous swap.
Around \dev{दवा}{davā} (medicine) the list absorbs \dev{दवाई}{davāī} and \dev{दवाएं}{davāen}, which
mean the same thing, but refuses \dev{दावा}{dāvā} (claim) and \dev{दबा}{dabā} (pressed), which only
sound the same. \dev{पत्ता}{pattā} (leaf) absorbs \dev{पत्ते}{patte} but refuses
\dev{पता}{patā} (address) (Figure~\ref{fig:families}). A corrector that rewrote every near-homophone
to its most common neighbour would turn medicine into claim and leaf into address. Those refusals
are what stop it.

The different-meaning pairs are not spread evenly across the categories, and where they pile up is
in \S\ref{sec:res-correction}.

Four weights are shared by the repair stage and by AWWER~\citep[\S3.3.2]{chandrashekar2026asr}:
weight~4 for core agriculture (crops, pests, practices), weight~3 for strongly related words (soil,
weather, timing), weight~2 for indirectly related words (quantities, locations), and weight~1 for
anything not in the list. 

\begin{table}[ht]
\centering\tabsize
\caption{How the sound-alike pairs were labelled, with an example of each\llmj.}
\label{tab:pairlabels}
\begin{tabularx}{\textwidth}{@{}l r r r l L@{}}
\toprule
Label & Pairs & Share & Conf. & Example & What happens to it \\
\midrule
Different meaning & 10,335 & \flag{74\%} & 0.80 & \dev{खाद}{khād} / \dev{खाद्य}{khādya} & Flagged with a cue, never rewritten \\
Same word, two spellings & 2,287 & 16\% & 0.85 & \dev{मकई}{makaī} / \dev{मकाई}{makāī} & May be folded to the main spelling \\
Same meaning, two words & 647 & 5\% & 0.69 & \dev{मक्का}{makkā} / \dev{मकई}{makaī} & May be folded to one concept \\
Unrelated & 635 & 5\% & & \dev{बोवाई}{bovāī} / \dev{मोबाइल}{mobāil} & Dropped from the index \\
\bottomrule
\end{tabularx}
\end{table}

``Conf.'' is the mean confidence the judge model attached to that label (Claude Sonnet 5). The
same-meaning class has the lowest confidence, which is why step 7 of Appendix~\ref{app:lexicon}
checks that class again before folding anything.

\subsubsection{Correction Rule}
\label{sec:m3-algo}
\begin{figure}[ht]
\centering
\includegraphics[width=0.84\textwidth]{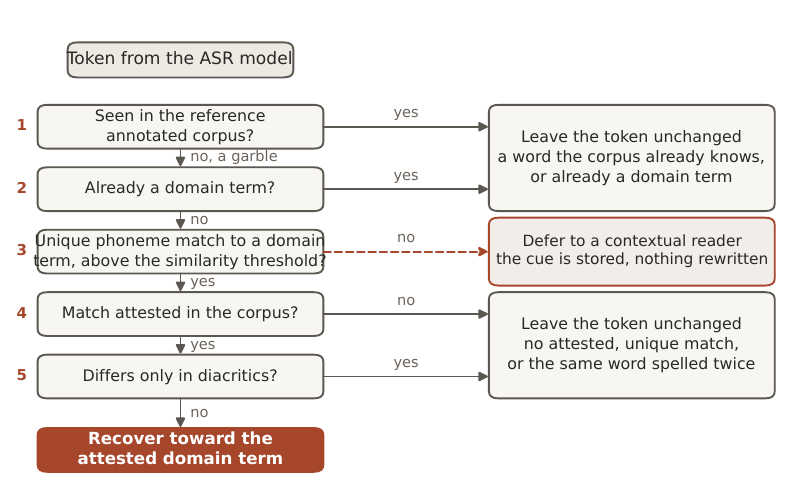}
\caption{The repair decision path. Structure only; measured effects are in
\S\ref{sec:res-correction}.}
\label{fig:corrflow}
\end{figure}

The rule has one job: repair a farming word the ASR model mangled, and leave every other word
alone. It is built to be careful rather than thorough. The reference keeps whatever the farmer
actually said and is never tidied up (\S\ref{sec:method-annotation}), so changing a word that was
already right can only lose points. Earlier work on fixing names in transcripts uses spelling
distance, sound similarity and sentence context
together~\citep{garg2020entitycorrector,ma2024asrcorrection}. This stage uses the first two and
leaves anything that needs context alone. Figure~\ref{fig:corrflow} is the path a word
takes.\footnote{The open-source Handy dictation tool~\citep{handy2025} was the reference for putting
a sound-matching step here. Handy uses Soundex, which does not work on Devanagari, so the sound
layer here was built for Indic scripts instead.}

\begin{itemize}
  \item \textbf{Which words are looked at.} Only a word real farmers almost never say. If the word
    appears in the query corpus, people use it, so the stage leaves it alone. Words the ASR model
    effectively made up are the only candidates.
  \item \textbf{How two words are compared.} By the sounds they are spoken with, not the letters
    they are written with, using a rule that also drops the silent vowel at the end of a Devanagari
    word. Swapping a soft sound for a hard one is a small difference; moving where in the mouth a
    sound is made is a bigger one~\citep{levenshtein1966binary,kondrak2000alignment}.
  \item \textbf{What a word may be changed into.} Only a farming term the ASR model is already known
    to produce, and only when that one term is a clear winner. If two terms are about equally close,
    nothing changes. 
  \item \textbf{Two spellings of one word.} Devanagari writes some words two ways, a dot where the
    other spelling puts a small curve. Those are the same word, not a mistake, so they are skipped.
    Without this the stage would keep rewriting good spelling and dialect variants.
  \item \textbf{Sound-alike words that mean different things are left alone.} The expensive cases,
    \dev{दवा}{davā} against \dev{दावा}{dāvā} and \dev{बीज}{bīj} against \dev{बीच}{bīch}, cannot be
    settled without reading the sentence, because the wrong word is a real word too. Each is flagged
    with the list's cue and passed on unchanged (\S\ref{sec:future}).
  \item \textbf{Why ties are refused.} A quarter of the list sounds exactly like something else in it
    (Appendix~\ref{app:lexicon}). Guessing between two farming words that sound the same is worse
    than leaving the word alone.
  \item \textbf{One sound system, three languages.} The sounds use one Indic scheme rather than the
    international phonetic alphabet, so the same matcher works for Telugu and Odia as soon as those
    word lists exist. Today the list is Devanagari and the gate is on the script of each word, so
    Telugu and Odia script passes through untouched and Devanagari written on a Telugu or Odia clip
    is repaired as Hindi (Appendix~\ref{app:terms}). Those two word lists are the next build
    (\S\ref{sec:future}).
  \item \textbf{Settings.} The frequency cut-offs, the closeness threshold and the sound costs are
    all configuration, not numbers buried in the code (P3).
\end{itemize}

\subsection{M4: Transcript Quality and Recovery}
\label{sec:m4}
M4 is specified and built but not measured, so it is presented here as design. No result in this
paper depends on it. What it is for is the gap of \S\ref{sec:fail-recovery}: no bad transcript
should reach the rest of the pipeline silently.

The stage rests on one property of the deployment. The model that writes the farmer's answer is told
which channel the question arrived on, text, voice or image. A model that knows the question was
spoken, and knows which words in the transcript of it are in doubt, can behave differently from one
handed plain text. That is what makes a recovery step possible without a second pass over the audio.

\begin{itemize}
  \item \textbf{A gate with two parts.} On top of a general confidence score, M4 flags a transcript
    when an important farming word is shaky: a crop, a chemical, a dose. The weights behind AWWER
    (\S\ref{sec:m3-lexicon}) already say which words those are, so the trigger is reasoned rather
    than one flat threshold. A wrong function word is survivable. A wrong pesticide is not.
  \item \textbf{What it can look at.} ASR model confidence where the provider gives it, the signal
    features M0 already computed, how many words M3 flagged as different-meaning pairs, and whether
    the repaired and unrepaired transcripts agree.
  \item \textbf{Recovery happens when the answer is written.} A flagged transcript goes to a
    higher-reasoning model at response generation, together with the channel and the list of farming
    words in doubt. That model can hedge, ask for the missing detail, or lean on the rest of the
    query instead of answering a word it should not trust. Recovery is therefore a property of
    response generation driven by the channel, not a second pass over the audio. A flagged clip can
    also be sent to another ASR provider or through a second repair pass.
  \item \textbf{This is where the judgements go that rules cannot make.} Every stage up to here is
    deterministic on purpose, and that is also its ceiling. A fixed rule cannot tell whether
    \dev{दवा}{davā} or \dev{दावा}{dāvā} was meant, whether a dose is plausible for the crop named in
    the same question, or whether an odd word is a garble or a local name. A language model makes
    those calls well and code does not make them at all. M4 is the one place in the design where
    such a call belongs, which is what keeps every measured stage free of a model call of its own
    (\S\ref{sec:alternatives}).
  \item \textbf{Thresholds and routing are configuration}, not code (P3), so a deployment sets how
    much it is willing to flag.
  \item \textbf{Asking the farmer, where the channel allows it.} An uncertain high-stakes word can
    be put back to the farmer to confirm instead of acted on. This sits outside the single-shot
    evaluation and affects none of the reported numbers.
\end{itemize}

We propose the stage and do not test it. What it is worth is how many bad clips it rescues against
how often it flags a clip that was already fine, and that needs its own experiment
(\S\ref{sec:future}). The closest thing this paper does measure is the weighted domain error of
\S\ref{sec:res-awwer}, because it counts exactly the words M4 is built to catch. Weighted domain
error is therefore the proxy for what this module would add: it says how many shaky farming words
are still in the transcript for M4 to act on, and it is the metric the M4 experiment should be read
against.

\section{Experiment Design}
\label{sec:method}

\subsection{Evaluation Data}
\label{sec:method-data}
Three datasets carry the work. Table~\ref{tab:datasets} lists them once, and we use the names after
that. Samples are spread across language, geography, crop, question type, noise level, number of
speakers, recording quality and how many farming words a clip holds.

\begin{table}[ht]
\centering\tabsize
\caption{The datasets used here. Scored counts per experiment are in each result table's caption.}
\label{tab:datasets}
\begin{tabularx}{\textwidth}{@{}l l l L@{}}
\toprule
Set & Scale & Languages & Role \\
\midrule
Published benchmark~\citep{chandrashekar2026asr} & 10,934 recordings evaluated & Hindi, Telugu, Odia & The earlier evidence this pipeline was built from; released as a dataset~\citep{digitalgreen2025dataset} \\
Reference annotated corpus & about 100,000 queries & Hindi & Mines the word list and gives corpus frequency for the repair stage (\S\ref{sec:m3}) \\
Human quality-checked corpus (IISc) & about 2,700 clips & Hindi, Telugu, Odia & The evaluation set: the diarizer, selection and the stage ladder; released with both references per clip~\citep{digitalgreen2026evalset} \\
\bottomrule
\end{tabularx}
\end{table}

A human quality-checked clip joins the evaluation set if it has audio and a farmer reference from
the human annotation. The ladder scores the clips where all four stages ran on all four ASR models. Both
counts are in the captions of Tables~\ref{tab:ladder} and~\ref{tab:awwer}. 

\subsection{Annotation Protocol}
\label{sec:method-annotation}
These are the guidelines that answer \S\ref{sec:fail-annotation}, run as a trial with the Indian
Institute of Science, Bangalore. Each clip gets six fields:

\begin{itemize}
  \item \textbf{Audio issues}, ticked from a fixed list: background noise, multiple speakers,
    overlapping speech, cross talk, radio or television, crowd noise, distorted audio, very low
    volume. More than one tick is normal.
  \item \textbf{Speaker count}, as a number.
  \item \textbf{Transcription with timestamps}, every speaker, as \texttt{(start -- end) S\#: text}.
    A pause over about two seconds or a change of subject starts a new segment. Where two people
    overlap, whoever is clearly audible is written down.
  \item \textbf{Main speaker}, the label of the person asking the farming question.
  \item \textbf{Farming words} used in the clip.
  \item \textbf{Category} for each of those words: crop, fertilizer, pesticide, disease, soil,
    irrigation, farming practice or weather.
\end{itemize}

Table~\ref{tab:annotrows} is one real clip with all six fields filled in. S1 is the helper holding
the phone, prompting the farmer. S2 is the farmer.

\begin{table}[ht]
\centering\tabsize
\caption{One clip as annotated, all six fields. Hindi, two speakers.}
\label{tab:annotrows}
\begin{tabularx}{\textwidth}{@{}l L@{}}
\toprule
Field & Value as annotated \\
\midrule
Audio issues & Background noise, multiple speakers \\
Speakers & 2 \\
Transcription & (00:00.8 -- 00:02.3) S1: \dev{भईया कौन तू क्वेशन पूछब?}{brother, which question are you going to ask} \newline
  (00:02.3 -- 00:04.3) S2: \dev{सरसो में कीड़ा लगता है, उसका बताइए?}{mustard is getting pests, tell me about that} \newline
  (00:04.8 -- 00:05.9) S1: unintelligible \\
Main speaker & S2 \\
Farming words & \dev{सरसों}{sarson}: crop; \dev{कीड़ा}{kīṛā}: pest \\
Reference for scoring & S2's turn only \\
\bottomrule
\end{tabularx}
\end{table}

Because people placed the timestamps, diarization error is measurable on this set directly. Because
the main speaker is labelled, so is speaker-selection accuracy.

Only the main speaker's turns become the reference, because the service has to answer the farmer's
question and not a bystander's. What was said is kept as said and never tidied up, so editing a word
the ASR model already got right can only lose points. That is the condition the repair stage is
built around (\S\ref{sec:m3-algo}).

\subsection{Text Normalization}
\label{sec:method-normalization}
Word error rate only means something if the reference and the transcript are cleaned up the same
way. In Indic scripts that takes care. The same word can differ by a dot under a letter, by a dot
above instead of a small curve, by a number written as digits or as words, or by annotation markup
in brackets. So scoring uses the benchmark's Hindi normalizer, ported unchanged, and runs the same
steps for Telugu and Odia rather than adding one-off rules per language.

\begin{itemize}
  \item \textbf{Hindi, ported from the benchmark~\citep{chandrashekar2026asr}.} Unicode NFC, fold the
    nukta so both ways of writing a letter collapse to one (\devplain{क़}~to~\devplain{क}), turn
    Devanagari digits into number words the way people say them in the Indian system of crore, lakh,
    thousand and hundred, strip bracketed annotation and punctuation, collapse whitespace.
  \item \textbf{Telugu and Odia, added here.} The same steps, each with its own digit map and
    number-word table, checked against AI4Bharat's \texttt{indic-numtowords} convention. Odia folds
    its own nukta, and both add the Indic danda to the punctuation list.
  \item \textbf{Corpus-level scoring.} WER is pooled: total edits over total reference words, as in
    the benchmark, rather than an average of per-clip rates. Each clip is normalized by its own
    language.
  \item \textbf{Shared sound space.} The repair stage (\S\ref{sec:m3-algo}) uses the same
    normalization plus schwa deletion by rule, which is what lets one word list and one matcher cover
    all three scripts as the other lists grow.
\end{itemize}

The Telugu and Odia number-word tables are built by rule and not yet reviewed by a native speaker,
so absolute WER in those two languages is comparable inside this paper rather than against outside
figures. Stage-to-stage
and model-to-model differences are unaffected, because the same normalizer runs on both sides.

\subsection{Pipeline Configurations}
\label{sec:method-configs}
The main protocol is a ladder. Each stage keeps everything before it and adds one thing
(Table~\ref{tab:configs}). Table~\ref{tab:modulekeys} names the modules those columns stand for.

\begin{enumerate}
  \item \textbf{Everything else is held still.} Same clips, same ASR model, same reference at every
    step. The only thing that changes is the audio or the text handed to the ASR model.
  \item \textbf{Selection sits with diarization in S3}, because finding the turns and picking the
    farmer are one operation.
  \item \textbf{Repair stands alone in S4}, so its effect is read on its own.
  \item \textbf{Selection and slicing read the audio the cleaning gate passes on}, so the ladder
    measures the stages in the order a deployment runs them.
  \item \textbf{M4 is measured on its own terms.} What it is worth is how many bad clips it
    rescues, not accuracy on the clips it accepts, so it needs its own experiment
    (\S\ref{sec:future}) rather than a rung on this ladder.
\end{enumerate}

\begin{table}[ht]
\centering\tabsize
\caption{The five modules the ladder columns stand for.}
\label{tab:modulekeys}
\begin{tabular}{@{}ll@{}}
\toprule
Module & What it does \\
\midrule
M0 & Audio analysis and gated cleaning (\S\ref{sec:m0}) \\
M1 & Speaker detection and selection (\S\ref{sec:m1}) \\
M2 & ASR model, unchanged and replaceable (\S\ref{sec:m2}) \\
M3 & Domain-aware correction against the farming word list (\S\ref{sec:m3}) \\
M4 & Transcript quality and recovery, specified in \S\ref{sec:m4} \\
\bottomrule
\end{tabular}
\end{table}

\begin{table}[ht]
\centering\tabsize
\caption{The four ladder stages and what each hands the ASR model. \tick{} on, \cross{} off.}
\label{tab:configs}
\begin{tabularx}{\textwidth}{@{}l L c c c c@{}}
\toprule
Stage & Input to the ASR model & M0 & M1 & M2 & M3 \\
\midrule
S1 Baseline & Raw audio, no cleaning & \cross & \cross & \tick & \cross \\
S2 Enhancement & Audio after gated cleaning (DFN3, 6~dB) & \tick & \cross & \tick & \cross \\
S3 Diarization and selection & The farmer's turns only, found by the fine-tuned segmenter and speaker selection & \tick & \tick & \tick & \cross \\
S4 Domain correction & The farmer's turns, then repair against the farming word list (\S\ref{sec:m3-algo}) & \tick & \tick & \tick & \tick \\
\bottomrule
\end{tabularx}
\end{table}

The ladder is cumulative: each stage keeps the ones before it, so a stage's contribution is the step
it adds. S1 is Baseline 0 of \S\ref{sec:baselines}. What a module adds on its own comes from the study
built for it: cleaning from the S1 to S2
comparison on the clips where it fired, sliced by noise tier (\S\ref{sec:res-denoise}); selection
from the diarizer comparison and the selector pilot (\S\ref{sec:res-diar}); repair from the
per-edit audit (\S\ref{sec:res-correction}).

\subsection{Metrics}
\label{sec:method-metrics}
We keep two kinds of error apart all the way through. Did the transcript match the reference, and
did the farmer's question survive. One example shows why both are needed: when \dev{खाद}{khād},
fertilizer, comes back as \dev{खाद्य}{khādya}, food, WER charges one word, the same price as a
dropped function word, while the farming-term rate charges the word the whole question was about
(Table~\ref{tab:confusions}). So a weighted rate and a farming-term rate are always reported next to
WER, and never one of them alone (Table~\ref{tab:metrics}).

\begin{table}[ht]
\centering\tabsize
\caption{What each metric measures and where it runs.}
\label{tab:metrics}
\begin{tabularx}{\textwidth}{@{}l L l@{}}
\toprule
Metric & What it measures & Where it runs \\
\midrule
Corpus WER & Wrong words against the farmer reference, pooled over the corpus & Every rung, and M0, M2, M3 \\
Farmer-query F1 & Recall against precision on the farmer reference, per clip & Every rung, and M0, M2, M3 \\
AWWER & WER with every error charged the importance of its word (\S\ref{sec:method-awwer}) & M3 and the ladder \\
Farming-term rates & Whether the crop, pest or chemical word survived, by word-list lookup & M3 and the ladder \\
DER & Missed speech, false alarm and speaker confusion, at a 0.25~s collar & M1 \\
cpWER & Error after each speaker is matched to a reference speaker~\citep{watanabe2020chime6} & M1 \\
Main-speaker accuracy & How often selection picks the labelled main speaker & M1 \\
Latency and cost & What each stage adds per query, and what it takes to deploy & Every stage \\
Answer quality & Out of scope: it rests on the model that reads the transcript & Not measured \\
\bottomrule
\end{tabularx}
\end{table}

WER reads a removed bystander the same as a removed farmer, and farmer-query F1 is what separates
them (Appendix~\ref{app:terms}). Farming terms are extracted with spelling variants folded on both
sides, so a variant spelling counts as a match instead of looking like a rescue.

Two markers appear throughout. A value measured on a sample rather than the full evaluation set
carries \pilot{}, with its sample size where it is quoted. A value a language-model classifier
produced, rather than code, carries \llmj{} and keeps that marker until people confirm it.

\subsection{Error-Cost-Aware Evaluation}
\label{sec:method-awwer}
Word error rate counts the insertions, deletions and substitutions needed to turn the transcript
into the reference, divided by the number of reference words. Every word costs the same, so a lost
crop name and a lost function word are charged alike. AWWER weights the errors by how much they
matter instead, using the four tiers of \S\ref{sec:m3-lexicon}:
\begin{equation}
\mathrm{AWWER} = \frac{\sum_{i \in \text{errors}} w_i}{\sum_{j \in \text{reference}} w_j},
\end{equation}
where $w$ is the weight from word-list lookup and anything not in the list gets weight~1~\citep[\S3.3.2]{chandrashekar2026asr}.
\begin{itemize}
  \item \textbf{Fixed and mechanical.} Words are lined up against the reference and charged by
    weight, with no language-model call at scoring time, though the word list the weights come from
    was adjudicated by one\llmj{}.
  \item \textbf{A reference word replaced or dropped} is charged its own importance.
  \item \textbf{A word the ASR model added} is charged its own importance too, so an invented crop
    name costs and an invented function word does not.
  \item \textbf{Weights are assigned once.} A judge model weights each word list entry, code applies
    it after that, and the judge is never the same model as the ASR model being tested.
\end{itemize}

AWWER does not by itself close the gap in \S\ref{sec:fail-eval}. An ASR model can look good on
AWWER and still fail the meaning check, which is why the farming-term rate of
\S\ref{sec:method-metrics} is reported with it.

\subsection{Run Coverage}
\label{sec:method-runplan}
The ladder ran to completion: the full evaluation set, all four stages, all four ASR models, all
three languages.

\begin{itemize}
  \item \textbf{ASR models.} Four off-the-shelf systems from different families and deployment
    styles, run through the identical pipeline: Gemini 3.7
    Flash~\citep{google2026gemini37} and Sarvam saaras:v3~\citep{sarvam2026saaras} (cloud), Azure
    Speech~\citep{microsoft2026azurespeech} (cloud), and IndicConformer
    600M~\citep{ai4bharat2025indicconformer} (on-device CTC). The model-agnostic claim is measured
    on all four.
  \item \textbf{Languages.} All three tested together on the full set. Repair is Hindi first,
    because the word list and its sound rules are Devanagari. It is gated on the script of each word
    rather than the language of the clip, so it also edits the Devanagari an ASR model sometimes
    produces on a Telugu or Odia clip (Appendix~\ref{app:terms}).
  \item \textbf{What each ASR model brings.} Gemini 3.7 Flash is cloud and generative and shows the
    biggest multi-speaker gain, Sarvam saaras:v3 is cloud and Indic-specialized, Azure Speech runs
    with cleaning on in the ladder even though its own evidence argues for switching it off, and
    IndicConformer 600M is on-device CTC
    with no API cost. No cell of the ladder is missing.
\end{itemize}

\section{Results}
\label{sec:results}
The ladder of \S\ref{sec:method-configs} ran to completion on the full evaluation set and on all
four ASR models. Every value here is a measured corpus value, produced from the committed per-clip
result files. Values from a sample carry \pilot{} and values a judge model produced carry \llmj{}.

\subsection{The Stage Ladder, S1 to S4}
\label{sec:res-ladder}
\begin{figure}[ht]
\centering
\includegraphics[width=0.92\textwidth]{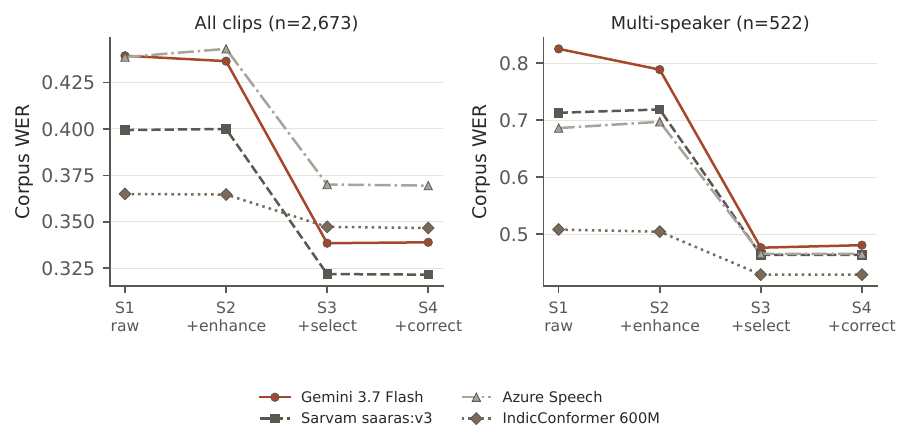}
\caption{Corpus WER across the four stages, one line per ASR model, all clips and multi-speaker
clips.}
\label{fig:ladder}
\end{figure}

The pipeline beats the plain ASR model on all four we tested, and the difference is
statistically significant each time. Against raw audio, the full pipeline cuts corpus WER by
\relcloudlo{} to \relcloudhi{} on the three cloud ASR models and by \relondevice{} on the
on-device model. On clips with more than one speaker it cuts error by \relmultilo{} to
\relmultihi{}. One stage carries almost all of that, and the on-device model gains the least
(Figure~\ref{fig:ladder}, Table~\ref{tab:ladder}).

\begin{table}[ht]
\centering\footnotesize
\caption{The stage ladder: corpus WER and farmer-query F1. Best value per row in green bold.}
\label{tab:ladder}
\begin{tabular}{@{}llrrrr@{}}
\toprule
Subset & ASR model & S1 & S2 & S3 & S4 \\
& & WER / F1 & WER / F1 & WER / F1 & WER / F1 \\
\midrule
\multirow{4}{*}{All ($n=2{,}673$)}
 & Gemini 3.7 Flash & 0.439 / 0.696 & 0.436 / 0.695 & \best{0.339} / 0.700 & 0.339 / \best{0.700} \\
 & Sarvam saaras:v3 & 0.399 / 0.720 & 0.400 / 0.720 & \best{0.322} / \best{0.729} & \best{0.322} / \best{0.729} \\
 & Azure Speech & 0.439 / 0.665 & 0.443 / 0.664 & 0.370 / \best{0.682} & \best{0.369} / \best{0.683} \\
 & IndicConformer 600M & 0.365 / 0.678 & 0.365 / 0.679 & 0.347 / \best{0.683} & \best{0.347} / \best{0.684} \\
\midrule
\multirow{4}{*}{Multi-speaker ($n=522$)}
 & Gemini 3.7 Flash & 0.825 / 0.617 & 0.789 / 0.612 & \best{0.476} / \best{0.656} & 0.480 / \best{0.656} \\
 & Sarvam saaras:v3 & 0.713 / 0.625 & 0.719 / 0.625 & \best{0.464} / \best{0.662} & \best{0.464} / \best{0.662} \\
 & Azure Speech & 0.686 / 0.591 & 0.697 / 0.590 & 0.466 / 0.642 & \best{0.465} / \best{0.643} \\
 & IndicConformer 600M & 0.508 / 0.619 & 0.504 / 0.624 & \best{0.429} / \best{0.641} & \best{0.429} / \best{0.641} \\
\midrule
\multirow{4}{*}{Single-speaker ($n=2{,}151$)}
 & Gemini 3.7 Flash & 0.337 / \best{0.715} & 0.343 / \best{0.715} & 0.302 / 0.710 & \best{0.301} / 0.711 \\
 & Sarvam saaras:v3 & 0.316 / 0.743 & 0.315 / 0.743 & \best{0.284} / \best{0.745} & \best{0.284} / \best{0.745} \\
 & Azure Speech & 0.373 / 0.683 & 0.376 / 0.682 & 0.345 / \best{0.692} & \best{0.344} / \best{0.692} \\
 & IndicConformer 600M & 0.327 / 0.692 & 0.328 / 0.693 & 0.326 / 0.693 & \best{0.325} / \best{0.694} \\
\bottomrule
\end{tabular}
\end{table}

Three things hold for all four ASR models.

\begin{itemize}
  \item \textbf{Speaker selection carries the pipeline.} It is the only stage with a big, reliable
    drop in error, and the drop is biggest where several people speak
    (Table~\ref{tab:ladderci}). On multi-speaker clips it removes between a sixth and nearly half of
    the word error. Farmer-query F1 goes up at the same time, so the gain is not just words being
    deleted.
  \item \textbf{Cleaning pays on the noisy tail.} At corpus level its effect sits inside the
    interval on three of the four ASR models, and on Azure it is a small reliable increase
    (Table~\ref{tab:ladderci}). The gain is in the high-noise clips and depends on the model
    (\S\ref{sec:res-denoise}), which is what the noise gate and the per-model switch are for.
  \item \textbf{Repair is conservative, and its value is in farming terms.} S3 to S4 is a small
    reliable drop on Azure and IndicConformer and a wash on Gemini and Sarvam, so it runs without a
    gate; on Gemini's multi-speaker clips it is a small reliable increase. What it is for is
    recovering garbled farming words, which it does on all four
    (\S\ref{sec:res-correction}, Appendix~\ref{app:terms}).
\end{itemize}

\begin{table}[ht]
\centering\footnotesize
\caption{Change in corpus WER per stage, paired clip bootstrap.}
\label{tab:ladderci}
\vspace{-2pt}
{\footnotesize\noindent Negative is better; * marks significance. Every S1 to S4 change is
significant.\par}
\vspace{2pt}
\begin{minipage}[t]{0.49\textwidth}
\centering
\textit{All clips}\\[2pt]
\begin{tabular}{@{}lrrr@{}}
\toprule
ASR model & S1$\to$S2 & S2$\to$S3 & S3$\to$S4 \\
\midrule
Gemini 3.7 Flash & $-$0.003 & \best{$-$0.098*} & $+$0.001 \\
Sarvam saaras:v3 & $+$0.001 & \best{$-$0.078*} & 0.000 \\
Azure Speech & $+$0.004* & \best{$-$0.073*} & \best{$-$0.001*} \\
IndicConformer 600M & 0.000 & \best{$-$0.017*} & \best{$-$0.001*} \\
\bottomrule
\end{tabular}
\end{minipage}\hfill\begin{minipage}[t]{0.49\textwidth}
\centering
\textit{Multi-speaker clips}\\[2pt]
\begin{tabular}{@{}lrrr@{}}
\toprule
ASR model & S1$\to$S2 & S2$\to$S3 & S3$\to$S4 \\
\midrule
Gemini 3.7 Flash & $-$0.036 & \best{$-$0.313*} & $+$0.004* \\
Sarvam saaras:v3 & $+$0.006 & \best{$-$0.255*} & 0.000 \\
Azure Speech & $+$0.011 & \best{$-$0.232*} & $-$0.001 \\
IndicConformer 600M & $-$0.004 & \best{$-$0.075*} & 0.000 \\
\bottomrule
\end{tabular}
\end{minipage}
\end{table}

\subsection{AWWER}
\label{sec:res-awwer}
WER charges the same for a lost crop name as for a lost function word: a dropped \dev{दवा}{davā},
the pesticide, costs exactly what a dropped postposition costs. AWWER exists to fix that
(\S\ref{sec:method-awwer}). Under AWWER the ladder tells the same story as under WER: selection
lowers weighted error on every ASR model, and repair leaves it where selection put it
(Table~\ref{tab:awwer}). The weights come from a Devanagari word list, so they cover 88.6\% of Hindi
reference words and charge Telugu and Odia words flat until those lists are built
(\S\ref{sec:future}). The Hindi column is the weighted view. AWWER is below WER everywhere in that
table. That says something about the
errors, not about the metric: it means the remaining errors sit on words that matter less than the
average reference word, so more of them are function words than crop, pest or chemical names.
Because an invented word is now charged its own importance instead of a flat~1, AWWER can be
compared across stages and across ASR models that differ in how much they write down.

\begin{table}[ht]
\centering\footnotesize
\caption{AWWER / WER pairs across the ladder, lexicon weights covering Hindi only. Hindi $n=1{,}130$, all languages $n=2{,}677$.}
\label{tab:awwer}
\begin{tabular}{@{}lrrrrrr@{}}
\toprule
& \multicolumn{3}{c}{Hindi, AWWER / WER} & \multicolumn{3}{c}{All languages, AWWER / WER} \\
\cmidrule(lr){2-4}\cmidrule(lr){5-7}
ASR model & S1 & S3 & S4 & S1 & S3 & S4 \\
\midrule
Gemini 3.7 Flash & 0.319 / 0.395 & \best{0.224} / 0.266 & \best{0.224} / 0.266 & 0.361 / 0.439 & \best{0.272} / 0.339 & \best{0.273} / 0.339 \\
Sarvam saaras:v3 & 0.330 / 0.384 & \best{0.255} / 0.283 & \best{0.255} / 0.282 & 0.351 / 0.399 & \best{0.282} / 0.322 & \best{0.282} / 0.322 \\
Azure Speech & 0.349 / 0.416 & 0.272 / 0.308 & \best{0.271} / 0.307 & 0.377 / 0.439 & 0.312 / 0.370 & \best{0.312} / 0.369 \\
IndicConformer 600M & 0.278 / 0.317 & \best{0.259} / 0.287 & \best{0.259} / 0.286 & 0.312 / 0.365 & \best{0.297} / 0.347 & \best{0.296} / 0.347 \\
\bottomrule
\end{tabular}
\end{table}

AWWER weights the errors but still scores the whole transcript. The narrower question, whether the
farmer's own crop, pest or chemical word survived, is measured directly in
Appendix~\ref{app:terms}: selection trades a little farming-term recall for more precision, and
repair raises that recall on every ASR model.

\subsection{S2 Enhancement}
\label{sec:res-denoise}
\begin{figure}[ht]
\centering
\includegraphics[width=0.80\textwidth]{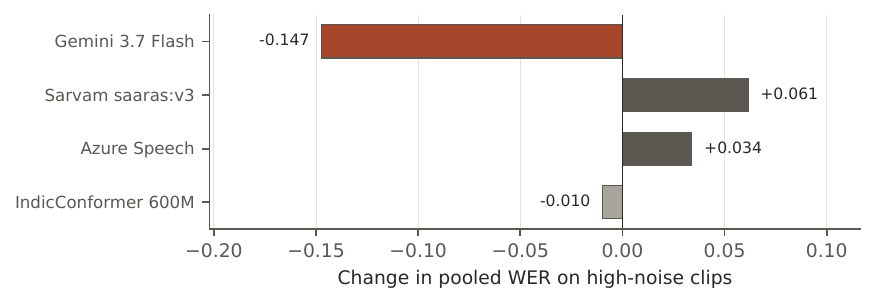}
\caption{Change in pooled WER from gated cleaning on high-noise clips, per ASR model. Negative is
better.}
\label{fig:denoiserec}
\end{figure}

Cleaning depends on the ASR model, and that dependence is the finding. On the high-noise clips it
lowers pooled WER for the generative cloud ASR model by 0.147, does almost nothing for the
on-device model at $-$0.010, and makes things worse for Sarvam ($+$0.061) and Azure ($+$0.034)
(Figure~\ref{fig:denoiserec}). Two readings are both true, and they answer different questions.
Pooled over the noisy clips, Gemini's gain is real and big enough to switch the stage on for that
ASR model. Clip by clip, the median change is zero in every noise band for every ASR model, and
about as many clips get worse as get better (Table~\ref{tab:denoisefull}), so the pooled gain comes
from a minority of clips rather than a general shift.

\begin{table}[ht]
\centering\tabsize
\caption{Gated cleaning on the full evaluation set, per ASR model. Negative is better.}
\label{tab:denoisefull}
\begin{tabularx}{\textwidth}{@{}>{\raggedright\arraybackslash}p{3.8cm} r l r l@{}}
\toprule
& \multicolumn{2}{c}{High noise} & \multicolumn{2}{c}{Medium noise} \\
\cmidrule(l){2-3}\cmidrule(l){4-5}
ASR model & Pooled & Better / worse / same & Pooled & Better / worse / same \\
\midrule
Gemini 3.7 Flash & \best{$-$0.147} & 28 / 20 / 46 & $+$0.008 & 158 / 164 / 231 \\
Sarvam saaras:v3 & $+$0.061 & 15 / 12 / 67 & $-$0.007 & 64 / 45 / 444 \\
Azure Speech & $+$0.034 & 20 / 18 / 56 & $+$0.016 & 93 / 110 / 350 \\
IndicConformer 600M & $-$0.010 & 15 / 11 / 68 & $+$0.000 & 77 / 73 / 403 \\
\bottomrule
\end{tabularx}
\end{table}

``Pooled'' is the change in corpus WER. The middle per-clip change is zero in every cell, which is
the point of the last column: cleaning leaves most clips alone and moves a few a long way.

The default setting costs error on the tier it should help, so 6~dB is the recommended
cap\pilot{} (200 clips).

So the gate is doing its job. Cleaning fires only above the noise threshold and only for an ASR
model whose own evidence supports it. It helps the model it helps and leaves the other two within a
thousandth of where they started at corpus level. Azure, where cleaning costs a reliable 0.004, is
the case the gate exists to switch off.

\subsection{S3 Diarization and Selection}
\label{sec:res-diar}
The part that makes S3 work is the diarizer. Nine systems were scored, seven open and two cloud, and
none of them split this audio cleanly, which is why one component was fine-tuned. In the scored
ladder selection runs on every clip, and the gate a deployment uses to skip the single-speaker
majority is measured on its own below. The diarizer values here are on a frozen held-out split,
every system against the same reference at the same 0.25~s collar.

\begin{figure}[ht]
\centering
\includegraphics[width=0.72\textwidth]{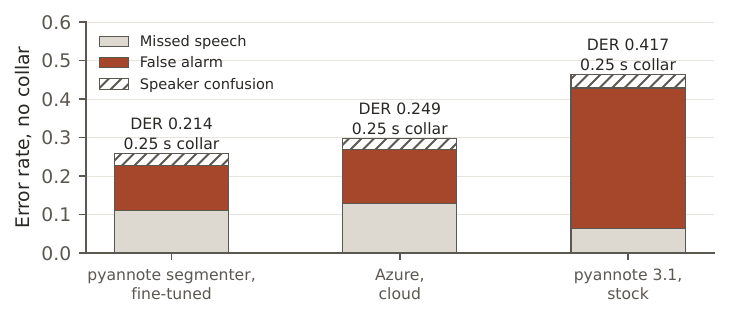}
\caption{Diarization error by component on the held-out split. Bars carry no collar, the annotation
does.}
\label{fig:diarder}
\end{figure}

\begin{figure}[ht]
\centering
\includegraphics[width=0.74\textwidth]{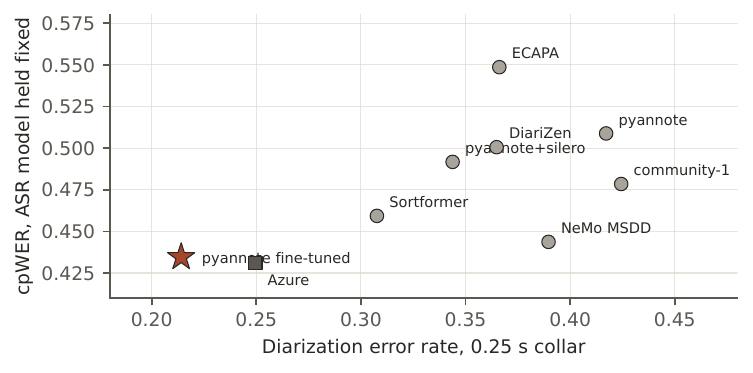}
\caption{Diarization error against cpWER with the ASR model held fixed, on the held-out split.
Bottom left is better.}
\label{fig:dercpwer}
\end{figure}

\begin{itemize}
  \item \textbf{The fine-tuned segmenter is the most accurate.} It reaches DER 0.214, against 0.249
    for the best cloud diarizer and 0.417 for stock pyannote (Figure~\ref{fig:diarder}).
  \item \textbf{What fell is false alarm}, which is speech heard in background noise. It drops from
    0.364 of stock pyannote's error to 0.118, paid for with missed speech rising from 0.065 to
    0.111. Components are scored without a collar, so they sum above the collared totals. The trade
    suits the common case of one bystander, and \S\ref{sec:disc-failures} says what a deployment
    with more group calls changes.
  \item \textbf{A lower DER does not mean a better transcript.} With the ASR model held fixed on
    each diarizer's segments, DER and cpWER move together only loosely (Pearson $r=0.60$, Spearman
    $\rho=0.57$). NeMo MSDD is mid-pack on DER and second-best on cpWER, ECAPA is mid-pack on DER
    and worst on cpWER, and the fine-tuned segmenter's cpWER of 0.435 is a near tie with Azure's
    0.431 despite a clear DER lead (Figure~\ref{fig:dercpwer}).
  \item \textbf{So diarizers are ranked by cpWER} with a fixed ASR model. Hearing too much speech
    costs DER seconds but barely costs words, which is why the segmenter was fine-tuned instead of
    taking whichever model had the lowest DER.
  \item \textbf{What selection removes} is the person in Table~\ref{tab:annotrows} who holds the
    phone and prompts the farmer. Text-only rules find the labelled main speaker well across the
    corpus and clearly worse on the multi-speaker clips that matter (Table~\ref{tab:selection}).
  \item \textbf{Ungated, selection costs the farmer's own words.} On single-speaker clips there is
    no bystander to remove, and slicing to turns cuts the farmer at the edges: a farmer who starts
    talking before the segmenter marks the turn loses those opening words. On 90 single-speaker
    pilot clips WER falls from 0.455 to 0.374 while farmer-query F1 falls from 0.712 to
    0.667\pilot{}.
  \item \textbf{The gate that ships needs no labels.} Run selection only when a stock diarizer sees
    a second speaker hold the floor for at least a second. On 149 pilot clips that gate agrees with
    the annotated speaker count 76\% of the time, lifts all-clip F1 from 0.678 to 0.686 and lowers
    WER from 0.630 to 0.422\pilot{}. On a bare speaker count, with no duration floor, it is 69\%
    accurate and lands below doing nothing at all.
\end{itemize}

\begin{table}[ht]
\centering\tabsize
\caption{Text-only speaker-selection rules against the annotated main-speaker label,
$n=2{,}716$ clips. Best per column in green bold.}
\label{tab:selection}
\begin{tabular}{@{}lrr@{}}
\toprule
Rule & All clips & Multi-speaker clips \\
\midrule
Longest single turn & \best{0.949} & \best{0.770} \\
Most total speech & 0.944 & 0.747 \\
Most turns & 0.938 & 0.713 \\
Earliest speaker & 0.882 & 0.428 \\
\midrule
Random choice & 0.879 & 0.413 \\
\bottomrule
\end{tabular}
\end{table}

\subsection{S4 Domain Correction}
\label{sec:res-correction}
The repair stage is deliberately narrow, and Table~\ref{tab:corraudit} measures how narrow.

\begin{itemize}
  \item \textbf{It touches few clips.} Across all four ASR models it changes 194 clips: 37 get
    better on corpus WER, 17 get worse, and 140 do not move.
  \item \textbf{The kind of edit it makes.} On a wheat query, \dev{खरपतवा}{kharpatvā}, which is not
    a word, goes back to \dev{खरपतवार}{kharpatvār}, weed, which is what the farmer's own reference
    says. A word the corpus does use is left alone, whatever it sounds like.
  \item \textbf{The harm is small, and worth reporting.} Those 17 are 0.16\% of all
    clip-ASR model pairs and 8.8\% of the clips the stage touched. Touching only words the corpus
    has never seen is what keeps that rate this low without a gate, and wrongly rewriting a crop or
    chemical name is worse than not repairing at all.
  \item \textbf{The harm is not spread evenly.} 14 of Gemini's 41 edits raise WER, against none of
    Azure's 41. Gemini writes down more of what it hears, so a word the corpus has never seen is
    more often a real rare word than a garble.
  \item \textbf{It does most on the on-device model.} IndicConformer garbles more farming words
    than the cloud systems do, so one word list rescues more of them: the stage touches 70 of its
    clips against 41 or 42 for the others, and raises its farming-term recall
    (Appendix~\ref{app:terms}). That is the ASR model a deployment cannot simply swap for a
    stronger cloud API.
\end{itemize}

\begin{table}[ht]
\centering\tabsize
\caption{What the repair stage touches, per ASR model, on the 2{,}677 scored clips.}
\label{tab:corraudit}
\begin{tabular}{@{}lrrrr@{}}
\toprule
ASR model & Clips changed & WER improved & WER worsened & WER unchanged \\
\midrule
Gemini 3.7 Flash & 41 & 6 & 14 & 21 \\
Sarvam saaras:v3 & 42 & 8 & 2 & 32 \\
Azure Speech & 41 & 11 & 0 & 30 \\
IndicConformer 600M & 70 & 12 & 1 & 57 \\
\midrule
All four & 194 & 37 & 17 & 140 \\
\bottomrule
\end{tabular}
\end{table}

\begin{figure}[ht]
\centering
\includegraphics[width=\textwidth]{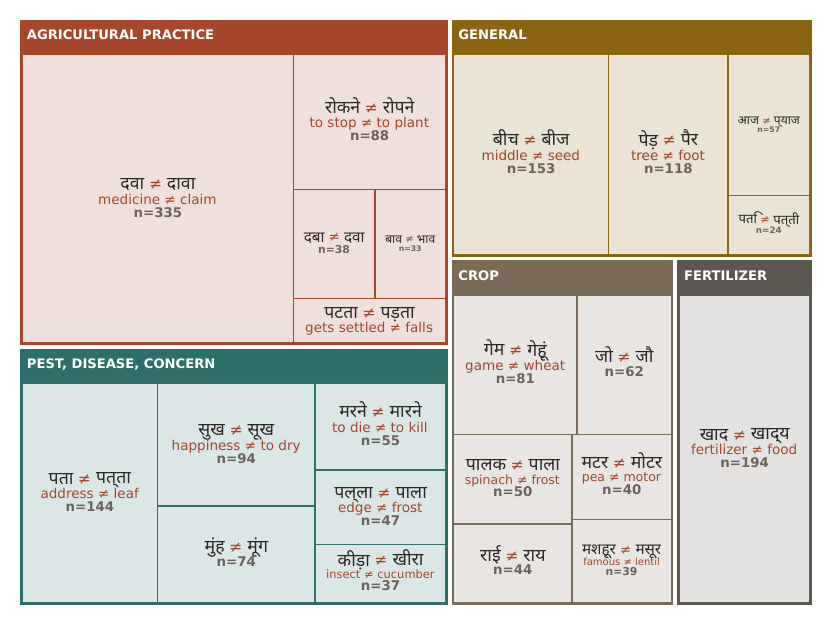}
\caption{Different-meaning confusions by category. Tile area is the observed count.}
\label{fig:confcat}
\end{figure}

What the stage has to leave alone is not spread evenly across the word list. Grouped by category,
the different-meaning pairs of \S\ref{sec:m3-lexicon} pile up in the categories the weighting scores
highest, farming practice and crops above all, because a word farmers ask about often is a word the
ASR model mishears often (Figure~\ref{fig:confcat}). Those are the pairs repair flags and never
rewrites, so the categories that matter most are also the ones where a contextual pass would buy the
most (\S\ref{sec:future}).

\subsection{Cost and Latency}
\label{sec:res-cost}
The pipeline adds a small, mostly CPU-bound cost around an unchanged ASR model
(Table~\ref{tab:cost}). No stage adds a second ASR call, and repair adds no model call at all.

\begin{table}[ht]
\centering\tabsize
\caption{What each stage adds in latency and cost. The ASR model's own bill does not change.}
\label{tab:cost}
\begin{tabularx}{\textwidth}{@{}l L L@{}}
\toprule
Stage & Latency & Cost \\
\midrule
M0 cleaning & Real-time factor about 0.02, roughly 113~ms median on a clip where it fires. The gate skips about 76\% of traffic as already clean, so about 27~ms per query on average & Cleaner is CPU only; the deployment gate is waveform arithmetic, and the runs here read the corpus noise tier \\
M1 diarization and selection & Real-time factor 0.019, estimated for the pyannote family on one A10G GPU & About \$0.02 per audio-hour self-hosted, that factor at full GPU use counting compute only, against \$0.18 per audio-hour published for the cloud service \\
M2 ASR model & Unchanged & Unchanged; no API cost at all with the on-device model \\
M3 repair & Text only, no extra ASR or model call & No model call \\
\bottomrule
\end{tabularx}
\end{table}

\section{Discussion}
\label{sec:discussion}

\subsection{Remaining Failure Modes}
\label{sec:disc-failures}
Four failures are left, each with a route out named against it. The places where a stage moves
error the wrong way are reported with that stage in \S\ref{sec:results}, and each is why the stage
is gated the way it is.

\begin{itemize}
  \item \textbf{Sound-alike words on clean audio.} When the wrong word is also a real word, only
    the sentence can tell you which was meant. The fixed rules leave these alone on purpose, for a
    reader that can use context (Table~\ref{tab:pairlabels}, \S\ref{sec:future}).
  \item \textbf{Clips with three or more speakers.} The fine-tuned segmenter accepts a little more
    missed speech in exchange for a large drop in false alarms, which is the right trade for the one
    bystander that most clips carry (\S\ref{sec:res-diar}). The threshold is configuration (P3), so
    a deployment that sees more group calls can move it back toward recall.
  \item \textbf{Words lost at turn edges.} Cutting the audio down to the farmer's turns costs a
    little of the farmer's own farming vocabulary (Appendix~\ref{app:terms}). The selection errors
    that remain are cases where a bystander simply talks longer than the farmer. The next lever is
    a content cue, since the question usually names the crop or the problem.
  \item \textbf{A dropped word cannot be repaired.} If the ASR model never wrote the farming word
    down, there is nothing to fix. Only ASR-side routes help there: cleaning, a vocabulary
    hint at decode time, or asking the ASR model for several candidates.
\end{itemize}

Odia is the hardest language for the generative ASR model. It has the most high-noise audio
(Figure~\ref{fig:noise}), the largest raw error on the farmer reference, and it is still the hardest
of the three after all four stages (Table~\ref{tab:ladderlang}).

A better transcript is not automatically a better answer. In a small pilot, enhanced and unenhanced
transcripts went through the whole advisory path and were compared by a blind judge with position
controlled; the transcript gain on high-noise clips did not change which answer was preferred,
because on the hardest clips both transcripts are wrong in different ways\pilot\llmj. None of these
failures is silent by design, which is why the quality gate (M4) sits in the architecture
(\S\ref{sec:m4}); measuring what it rescues is the next experiment (\S\ref{sec:future}).

\subsection{Generalizability}
\label{sec:disc-general}
\begin{itemize}
  \item \textbf{Works across ASR models.} The same pipeline, unchanged, improves all four
    ASR models significantly, and the gain is biggest where the audio is hardest and
    where the ASR model writes down the most bystander speech, without hurting the easy
    single-speaker case.
  \item \textbf{Works across languages.} Cleaning, diarization and selection listen to audio, so
    they behave the same in all three languages. Repair reaches Devanagari today, and because it is
    built on a sound alphabet shared across scripts the same matcher extends as the Telugu and Odia
    word lists are built (\S\ref{sec:future}).
  \item \textbf{Across geographies.} The next test is field audio at benchmark scale from another
    region (\S\ref{sec:future}).
  \item \textbf{Beyond farming.} The shape of the system does not depend on the domain: wrap an
    untouched ASR model, aim one specialist stage at each failure, and switch each stage on only
    where it helps. Only the word list is domain-specific, and the way it is built works for any
    other controlled vocabulary (Appendix~\ref{app:lexicon}).
\end{itemize}

\subsection{Engineering Implications}
\label{sec:disc-engineering}
\begin{itemize}
  \item Every stage is a separate part behind a shared interface, so a team can roll this out piece
    by piece. Add speaker selection first, because it is the biggest lever. Leave cleaning off for a
    ASR model it hurts. Swap the model without touching anything else.
  \item The extra cost is small and mostly CPU work (Table~\ref{tab:cost}), which matters for a
    live service where an accuracy gain that multiplies latency or the bill cannot ship.
  \item Provider independence is the practical payoff. The ASR model is the one swappable cost
    centre, and the stages around it carry the domain knowledge that would otherwise force a
    fine-tune, so a deployment can trade accuracy against cost without a redesign.
\end{itemize}

\section{Future Work}
\label{sec:future}
Each item below is scoped, and most are already specified. This is what the pipeline needs next.

\begin{itemize}
  \item \textbf{Measure M4.} How many bad clips the quality gate rescues, against how often it
    rescues a clip that was already fine.
  \item \textbf{A contextual pass for sound-alike words.} Let the model that reads the transcript
    settle the pairs the fixed rules only flag (Table~\ref{tab:pairlabels}).
  \item \textbf{Repair in Telugu and Odia.} Each needs a mined word list, labelled pairs and its own
    sound layer before S4 means anything.
  \item \textbf{More languages.} Amharic, Afaan Oromo, Swahili and Hausa next, each with its own word
    list and sound layer.
\end{itemize}

\section{Conclusion}
\label{sec:conclusion}
General-purpose ASR models transcribe noisy, multi-speaker, language-mixed farmer audio poorly, and the
words they lose are the crop, pest and chemical names that carry the question. Five cheap stages, one of them the untouched
ASR model, fix much of that: clean the audio when it needs it, split the speakers and keep the
farmer, recognize with the ASR model unchanged, repair garbled farming words against a checked word
list, and judge the transcript before it goes on. No ASR model was retrained, no provider was replaced, and no
large agent system was built, and no stage of the pipeline makes a language-model call of its own.

On human quality-checked field recordings in Hindi, Telugu and Odia it cuts word error rate by
\relcloudlo{} to \relcloudhi{} on the three cloud ASR models and by \relondevice{} on the on-device
model, and by \relmultilo{} to \relmultihi{} on multi-speaker audio. Every one of the four
ASR models improves significantly. That the same pipeline works across model families and across
cloud and on-device deployment is what makes the claim about ASR model independence real rather
than asserted. The big lever is isolating the farmer's own speech. Cleaning and repair are switched
on only where the evidence supports them, and that evidence came from measurement, not assumption.
One component is fine-tuned, the diarization segmenter, because nothing off the shelf worked across
these three languages. Every other stage is off the shelf behind a shared interface.

Two findings carry beyond this system. Cleaning up audio does not reliably help recognition, so a
stage that makes audio sound better has to be gated on what the ASR model does with it. And when the
reference keeps the speaker's own words, repair has to aim at precision rather than coverage. The
pipeline is released with its word list and its evaluation code, and its stages can be deployed and
swapped one at a time.

\section{Open-Source Release}
\label{sec:release}
Four artifacts are public now: the evaluation set behind every result in \S\ref{sec:results}, the
farming word list, the one fine-tuned checkpoint, and a demonstration of the pipeline on sample
clips. The pipeline and evaluation code is published with this paper. Table~\ref{tab:release} lists
each one with its licence.

\begin{table}[ht]
\centering\tabsize
\caption{The open-source release, with licence and availability per artifact.}
\label{tab:release}
\begin{tabularx}{\textwidth}{@{}>{\raggedright\arraybackslash}p{3.4cm} L l l@{}}
\toprule
Artifact & What it holds & Licence & Available \\
\midrule
Evaluation set~\citep{digitalgreen2026evalset} & Field audio with human transcripts in Hindi, Telugu and Odia. Every clip carries both the farmer reference used for scoring here and the full per-speaker transcription with turn timestamps, plus speaker count and audio issues & CC-BY-4.0 & Now \\
Farming word list~\citep{digitalgreen2026lexicon} & The weighted terms, the adjudicated sound-alike pairs with their cues, the candidate neighbours and the variant families of Appendix~\ref{app:lexicon} & CC-BY-4.0 & Now \\
Diarization segmenter~\citep{digitalgreen2026segmenter} & The fine-tuned checkpoint, which runs on CPU, with its training recipe and its held-out result & MIT & Now \\
Demonstration~\citep{digitalgreen2026demo} & The pipeline stage by stage on sample clips: raw, cleaned and farmer-only audio, with each transcript against the human reference & n/a & Now \\
Pipeline and evaluation code~\citep{digitalgreen2026code} & The stage implementations, the ladder runner, the diarizer comparison, the corrector, the significance testing, and the figure and table generators & n/a & With this paper \\
\bottomrule
\end{tabularx}
\end{table}

With the code, the evaluation set, the word list and the checkpoint, every value in the result tables
of \S\ref{sec:results} regenerates from the committed per-clip result files. The field audio the
segmenter trained on is not redistributed, so that artifact ships as weights and recipe rather than
as a training set. The earlier benchmark this work builds on is published
separately~\citep{digitalgreen2025dataset}.

\begingroup
\small
\setlength{\bibsep}{2pt plus 1pt}
\bibliographystyle{unsrtnat}
\bibliography{references}
\endgroup

\appendix
\small
\section{Agricultural Lexicon Construction}
\label{app:lexicon}
The word list of \S\ref{sec:m3-lexicon} is built in seven steps from roughly 100,000 real Hindi
farmer questions. Every step works from word frequency or from sound, so the same build runs for
another language once a comparable corpus exists. All counts are read from a committed file, and the
products are released~\citep{digitalgreen2026lexicon}. Figure~\ref{fig:lexfunnel} carries an example
through each stage.

\begin{enumerate}
  \item \textbf{Mine.} Pull candidate farming words and phrases out of the query corpus, ranked by
    frequency. Single words and two-to-three word phrases both count, so multi-word names such as
    fertilizer brands survive.
  \item \textbf{Clean.} Normalize Unicode (NFC, fold the nukta), strip annotation marks, drop
    non-farming and non-Devanagari junk. The 127 removals: run-on phrase 55, non-Devanagari 41,
    Latin script 21, annotation artifact 7, numeric 3. A further 161 rows fold into an existing
    entry, leaving 18,646.
  \item \textbf{Tier and weight.} Give each entry a tier (core 2,771, long tail 13,015, phrase
    2,860), a category (13; the largest are pest, disease or concern 4,672, practice 4,153, crop
    1,785, fertilizer or chemical 1,068) and a weight (4: 9,195 terms; 3: 7,498; 2: 1,953).
  \item \textbf{Phonemize.} Write each entry as the sounds it is spoken with, in one scheme shared
    across Indic scripts, dropping the silent final vowel and the rule-governed middle ones
    (\S\ref{sec:m3-algo}). A quarter of entries, 4,756 of 18,646 in 1,957 groups, sound like
    another entry, which is why the matcher refuses a tie.
  \item \textbf{Cluster.} For each of 2,563 headwords, collect the entries that sound close to it:
    the sets an ASR model could confuse.
  \item \textbf{Adjudicate.} Label every pair as the same word, the same meaning, a different
    meaning, or unrelated, with a cue on each different-meaning pair. Of 13,904 pairs: same word
    2,287, same meaning 647, different meaning 10,335, unrelated 635
    (Table~\ref{tab:pairlabels})\llmj{}.
  \item \textbf{Family and over-merge guard.} Group same-meaning variants into 398 families
    comprising 950 member words, mean size 2.39, then recheck every merge: 118 members change
    meaning and are pulled back out, leaving 2,934 pairs safe to fold and 10,335 context-only
    pairs.
\end{enumerate}

The pipeline uses two products: the 2,934 safe pairs, folded only when the ASR model produced a word
the corpus has never seen (\S\ref{sec:m3-algo}), and the 10,335 different-meaning pairs with their
cues, flagged and never rewritten. Step 7 is what makes the list safe: without it, folding one
variant family would have collapsed 118 distinct meanings, including the \dev{दवा}{davā} and
\dev{दावा}{dāvā} case of Figure~\ref{fig:families}. Sound-alike chains join 2,631 of the 2,771 core
words into one group, so the pairs are overlapping neighbourhoods, not clean partitions.

\clearpage
\section{Farming-Term and Per-Language Detail}
\label{app:terms}
Both views here repeat under a second metric what \S\ref{sec:res-ladder} already shows under WER.

\begin{itemize}
  \item \textbf{Selection loses a few farming words and gains more precision.} At S3 recall falls
    slightly, precision rises, and F1 rises on all four ASR models, sharply so on multi-speaker
    clips (Table~\ref{tab:agriterms}). The precision gain is the bystander's farming words leaving
    the transcript; the recall loss is the farmer's own words clipped at turn edges, the harm
    corpus WER hides and why selection is gated (\S\ref{sec:res-diar}).
  \item \textbf{Repair raises farming-term recall on every ASR model.} S3 to S4 moves recall up on
    all four, the stage doing the job it was built for, while its effect on corpus WER is at most a
    thousandth: it shows its worth on the metric it targets, not the one averaging over every word.
  \item \textbf{What F1 adds to WER.} WER charges a lost crop name the price of a lost function
    word, and reads a removed bystander the same as a removed farmer. F1 and the farming-term rate
    separate those cases.
\end{itemize}

\begin{table}[ht]
\centering\scriptsize\setlength{\tabcolsep}{3.5pt}
\caption{Farming-term recall / precision / F1 across the ladder, Hindi. Cells are recall /
precision / F1.}
\label{tab:agriterms}
\begin{tabular}{@{}llccc@{}}
\toprule
Subset & ASR model & S1 & S3 & S4 \\
\midrule
\multirow{4}{*}{\shortstack[l]{All Hindi\\($n=1{,}120$)}}
 & Gemini 3.7 Flash & 0.881 / 0.809 / 0.831 & 0.856 / 0.832 / 0.836 & 0.857 / 0.831 / 0.837 \\
 & Sarvam saaras:v3 & 0.832 / 0.826 / 0.816 & 0.824 / 0.859 / 0.831 & 0.825 / 0.856 / 0.830 \\
 & Azure Speech & 0.829 / 0.800 / 0.801 & 0.818 / 0.837 / 0.818 & 0.819 / 0.835 / 0.817 \\
 & IndicConformer 600M & 0.811 / 0.832 / 0.809 & 0.803 / 0.839 / 0.812 & 0.805 / 0.834 / 0.811 \\
\midrule
\multirow{4}{*}{\shortstack[l]{Multi-speaker\\($n=353$)}}
 & Gemini 3.7 Flash & 0.870 / 0.706 / 0.755 & 0.805 / 0.764 / 0.772 & 0.807 / 0.762 / 0.772 \\
 & Sarvam saaras:v3 & 0.801 / 0.734 / 0.744 & 0.777 / 0.809 / 0.779 & 0.778 / 0.804 / 0.776 \\
 & Azure Speech & 0.805 / 0.708 / 0.734 & 0.782 / 0.790 / 0.771 & 0.783 / 0.787 / 0.770 \\
 & IndicConformer 600M & 0.769 / 0.768 / 0.752 & 0.759 / 0.797 / 0.767 & 0.761 / 0.792 / 0.765 \\
\bottomrule
\end{tabular}
\end{table}

Repair works off a Devanagari word list and is gated on the script of each word, so Hindi carries
almost all of what it moves: Telugu passes through untouched, and Odia shifts only where an ASR
model wrote Devanagari (Table~\ref{tab:ladderlang}). Selection carries the gain in all three
languages.

\begin{table}[ht]
\centering\tabsize
\caption{Per-language ladder, Gemini 3.7 Flash. Repair edits Devanagari, so it also touches Odia. Best per
row in green bold.}
\label{tab:ladderlang}
\begin{tabular}{@{}lrrrrr@{}}
\toprule
Language & $n$ & S1 & S2 & S3 & S4 \\
& & WER / F1 & WER / F1 & WER / F1 & WER / F1 \\
\midrule
Hindi & 1{,}126 & 0.395 / 0.772 & 0.393 / 0.770 & \best{0.266} / 0.787 & \best{0.266} / \best{0.788} \\
Telugu & 996 & 0.460 / \best{0.651} & 0.457 / 0.650 & \best{0.394} / 0.647 & \best{0.394} / 0.647 \\
Odia & 551 & 0.540 / \best{0.622} & 0.532 / 0.621 & \best{0.459} / 0.616 & 0.461 / 0.617 \\
\bottomrule
\end{tabular}
\end{table}

\end{document}